\documentclass[manuscript,trackchanges]{aastex61}
\hypersetup{linkcolor=magenta,citecolor=blue,filecolor=cyan,urlcolor=magenta}

\received{May 21, 2017}
\revised{June 18, 2017}
\accepted{June 20, 2017}

\shorttitle{Long-term Periodicities of CVs with Synoptic Surveys}
\shortauthors{YANG, Michael T.C. et al.}

\begin{document}
\title{Long-term Periodicities of Cataclysmic Variables with Synoptic Surveys}

\author{Michael Ting-Chang Yang}
\affil{Graduate Institute of Astronomy, National Central University, Jhongli,
Taiwan}
\email{mky@astro.ncu.edu.tw}

\author{Yi Chou}
\affil{Graduate Institute of Astronomy, National Central University, Jhongli,
Taiwan}

\author{Chow-Choong Ngeow}
\affil{Graduate Institute of Astronomy, National Central University, Jhongli,
Taiwan}

\author{Chin-Ping Hu}
\affil{Department of Physics, The University of Hong Kong, Pokfulam Road, Hong Kong}

\author{Yi-Hao Su}
\affil{Graduate Institute of Astronomy, National Central University, Jhongli,
Taiwan}

\author{Thomas A. Prince}
\affil{Division of Physics, Mathematics and Astronomy, California Institute of Technology, Pasadena, CA 91125, USA}

\author{Shrinivas R. Kulkarni}
\affil{Division of Physics, Mathematics and Astronomy, California Institute of
Technology, Pasadena, CA 91125, USA}

\author{David Levitan}
\affil{Division of Physics, Mathematics and Astronomy, California Institute of
Technology, Pasadena, CA 91125, USA}

\author{Russ Laher}
\affil{Infrared Processing and Analysis Center, California Institute of Technology, Pasadena, CA
91125, USA}

\author{Jason Surace}
\affil{Spitzer Science Center, California Institute of Technology, Pasadena, CA
91125, USA}

\author{Andrew J. Drake}
\affil{California Institute of Technology, Pasadena, CA 91125, USA}

\author{Stanislav G. Djorgovski}
\affil{Division of Physics, Mathematics and Astronomy, California Institute of Technology, Pasadena, CA 91125, USA}

\author{Ashish A. Mahabal}
\affil{Division of Physics, Mathematics and Astronomy, California Institute of Technology, Pasadena, CA 91125, USA}

\author{Matthew J. Graham}
\affil{California Institute of Technology,  Pasadena, CA 91125, USA}

\author{Ciro Donalek}
\affil{Division of Physics, Mathematics and Astronomy, California Institute of Technology, Pasadena, CA 91125, USA}



\begin{abstract}
A systematic study on the long-term periodicities of known Galactic cataclysmic
variables (CVs) was conducted.  Among 1580 known CVs, 344 sources were matched
and extracted from the Palomar Transient Factory (PTF) data repository.  The
PTF light curves were combined with the Catalina Real-Time Transient Survey
(CRTS) light curves and analyzed.  Ten targets were found to exhibit long-term
periodic variability, which is not frequently observed in the CV systems. These
long-term variations are possibly caused by various mechanisms, such as the
precession of the accretion disk, hierarchical triple star system, magnetic
field change of the companion star, and other possible mechanisms.  We discuss
the possible mechanisms in this study. If the long-term period is less than
several tens of days, the disk precession period scenario is favored.  However,
the hierarchical triple star system or the variations in magnetic field
strengths are most likely the predominant mechanisms for longer periods.
\end{abstract}


\section{Introduction} \label{sec:intro}
A cataclysmic variable (CV) is an accreting binary system composed of a white
dwarf (WD) as primary and a low mass companion.  In general, when the companion
fills its Roche lobe, the mass flow stream passing through the inner Lagrangian
point ($L_1$) generates an accretion disk around a non-magnetic WD.  On the
other hand, only a truncated disk can be formed in an intermediate polar (DQ
Her type), a subclass of CVs with a highly magnetic WD. The WD in the polar (AM
Her type) system has an even higher magnetic field that can prevent the
formation of the accretion disk.  The variability on different time scales of
CV systems is caused by different mechanisms.  The orbital periods of CVs
typically range from 70 min to 24 h, which is strictly related to the binary
separation and mass ratio.  A census on the orbital period distribution of CVs
reveals a period gap of approximately $2-3$ h
\citep[e.g.,][]{1995CAS....28.....W}, which is explained by the evolution
scenarios of CVs.

The variation with a time scale longer than a day is typically called the
superorbital or long-term variation in CV systems. Long-term variation has been
detected in only a few CVs, and this has been neglected in subsequent research.
\citet{2004RMxAC..20..238K} studied 100 CVs using the structure function to
characterize the time scales of the long-term variabilities.  However, no
further results and the implications behind the long-term variabilities were
addressed in this study.  Various types of mechanisms were proposed to explain
the long-term variations of CVs.  For example, \citet{2010A&A...514A..30T}
discovered a long-term modulation with a period of $4.43 \pm 0.05$ d in
cataclysmic variable PX And through eclipse analysis, which was considered to
be the disk precession period that triggers the negative superhump in this
system.  On the other hand, from the analysis of eclipse time variations in
eclipsing binary DP Leo, a third body with an elliptical orbit and a period of
$P = 2.8 \pm 2.0$ yrs was found by \citet{2011A&A...526A..53B}.
\citet{2014AJ....147...10H} discovered a period of $\sim$ 25 d oscillations,
regarded as the result of accretion disk instability in V794 Aql during its
small outbursts.  \citet{1988Natur.336..129W} proposed that the cyclical
variations of the orbital periods (on the time scale of years to decades) for
some CVs are related to their quiescent magnitudes and outburst intervals, and
the variations were inferred as the effect of the solar-type magnetic cycle of
the companion.  \citet{2012IAUS..282...91K} discovered several polars
exhibiting long-term variability with a time scale of hundreds of days, likely
caused by the modulation of the mass-transfer rate owing to the magnetic cycles
in the companion stars.

Previous studies on the long-term periodicities of CVs are sporadic.  With the
help of recent large synoptic surveys, we are able to search for and further
characterize the long-term variations of the CVs systematically.  In
Section~\ref{sec:obs}, we introduce the synoptic survey projects we utilized,
and the corresponding intensive observations made using the Lulin
One-Meter-Telescope (LOT).  Descriptions of our analysis method for the
long-term periodicity of the sources are presented in Section~\ref{sec:analy}.
The possible mechanisms driving the long-term variability and further
implications are presented in Section~\ref{sec:mechanism}.  In
Section~\ref{sec:longterm}, we summarize the long-term periodicities from our
results and discuss some of the previous studies related to the sources. 

\section{CV Catalog, Synoptic Surveys and Observations} \label{sec:obs}
The CVs selected for this study are from the catalog by
\citet{2006yCat.5123....0D}.  The data used for this study are from two
surveys: the Palomar Transient Factory (PTF) and the Catalina Real-Time
Transient Survey (CRTS).  In addition, the LOT, a small telescope capable of
intensive observations, was utilized for finding the orbital periods of the
targets that were unknown before this study.

\subsection{CV Catalog for Source Matching}
The CV catalog produced by \citet{2006yCat.5123....0D} (hereafter Downes'
catalog) is taken as a reference catalog for our study \citep[see][for the
catalog description]{2001PASP..113..764D}. The latest version of the catalog
contains 1830 sources, including 1580 CVs and 250 non-CVs.  The catalog used
the General Catalogue of Variable Stars (GCVS) name as the source identifier.
However, some of the sources have no GCVS names, and so the constellation name
of the sources was adopted as the identifier.  In this study, if multiple
sources without GCVS names are in the same constellation, then the
constellation name with distinct numbers were adopted as the project name of
the source (e.g., UMa 01 for one of the CVs in the constellation Ursa Major).
We used the Downes' catalog for our matching process and then retrieved the
light curves of the matches.

\subsection{Palomar Transient Factory}
The PTF \citep{2009PASP..121.1395L,2009PASP..121.1334R} project began observing
in 2009.\footnote{\url{http://www.ptf.caltech.edu/}} The Samuel Oschin
Telescope, a 48-in Schmidt telescope with Mould R, SDSS g$^\prime$, and several
H-$\alpha$ filters was adopted for the survey.  The PTF and its successor
intermediate PTF (iPTF) projects were accomplished in March 2017.  A 7.9 square
degree field of view was achieved with the camera configuration of PTF and
iPTF.  The next generation of the PTF project, called the Zwicky Transient
Facility (ZTF), with its many upgrades in software and hardware, will be
operated in mid-2017.

\begin{figure}[ht!]
\plotone{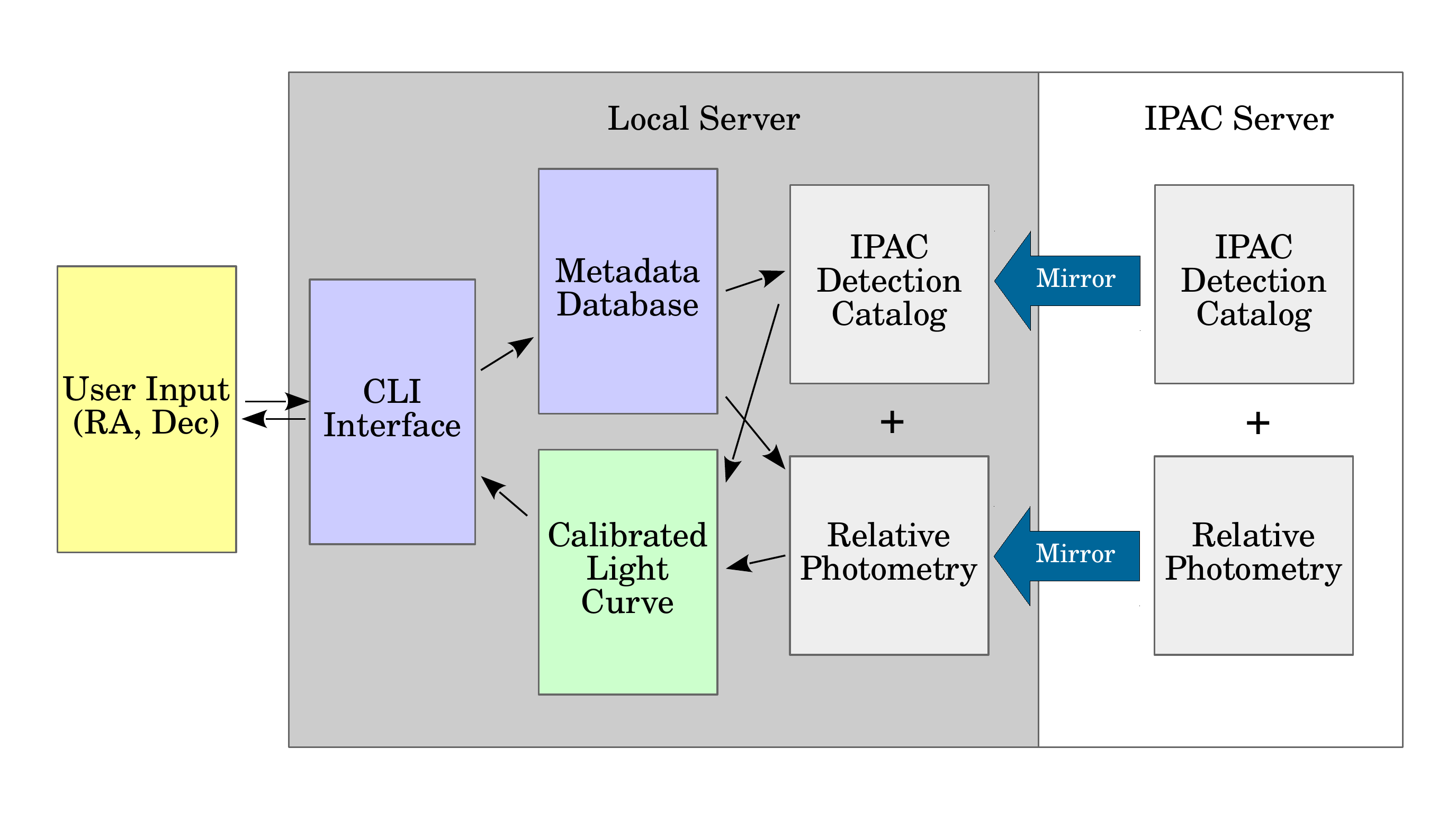}
\caption{Data flowchart of the light-curve-retrieval process for PTF.  IPAC
    data-processing pipeline \citep{2014PASP..126..674L} reduces the survey
    images, and generates IPAC detection catalogs and relative photometry for
    each frame (right panel).  Middle panel: programs, metadata database and
    mirrors of the processed data in our local server. The metadata database is
    generated to store the header information of the IPAC detection catalogs,
    and is used to speed up the data retrieval process.  A command line (CLI)
    interface was made for users to retrieve the calibrated light curve for a
    specific source.  The speed of the light-curve-retrieval process is
    increased by a significant factor in this implementation.
    \label{fig:flowchart}}
\end{figure}

\begin{figure}[ht!]
\plotone{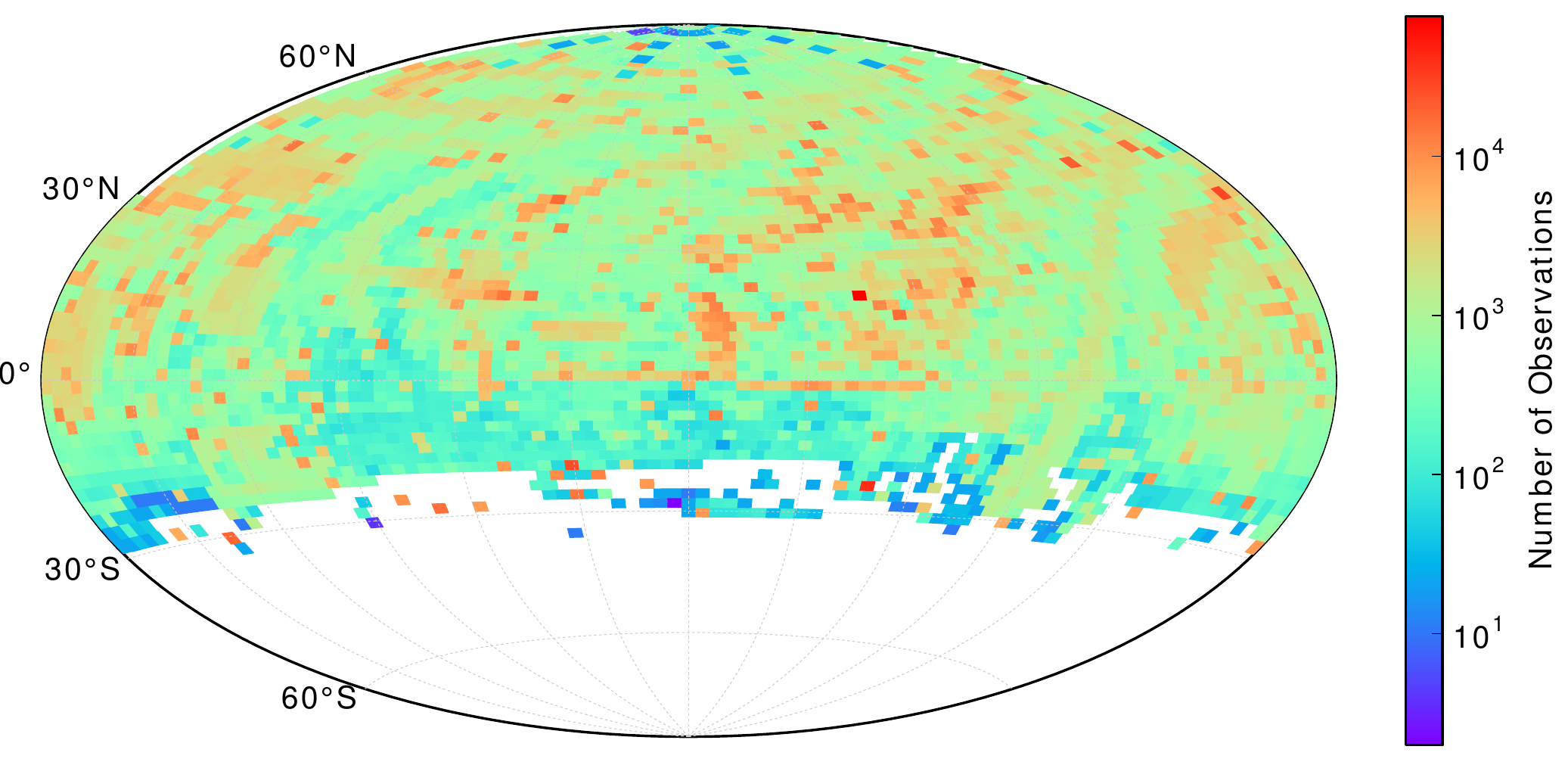}
\caption{Observational distribution of PTF survey. Each tile in the plot
    represents a PTF field, and is color-coded with the number of observations
    in the PTF photometric database. Only R band data are presented. The plot
    is in equatorial coordinates with the Hammer projection. Observations are
    absent in the Southern Equatorial Hemisphere (Decl.$<-30^\circ$) because of
    the observing limit of Palomar Mountain site. \label{fig:ptf_field}}
\end{figure}

The software pipelines were readied for the real-time discoveries of transients
\citep{2017PASP..129a4002M} and for the generation of source detections
\citep[IPAC pipeline,][]{2014PASP..126..674L}.  The IPAC pipeline was developed
by the Infrared Processing and Analysis Center
(IPAC)\footnote{\url{http://www.ipac.caltech.edu/}}, which reduced the images
and generated the detection catalogs on a frame basis.  The detection catalogs
are stored in the IRSA archive.\footnote{\url{http://irsa.ipac.caltech.edu}} We
extracted the data from the local copies of the full photometric catalogs in
IPAC. Metadata tables with information from the catalog headers were created
for accessing the photometric data quickly. The light curve of a specific
source could be retrieved out of the total of $\sim 24$ Terabytes within 2 min
via our data retrieval pipeline.\footnote{The IPAC web interface with faster
data retrieval process is currently online.} Figure~\ref{fig:flowchart} shows a
flowchart of the data-retrieval pipeline. A large area of the sky has been
observed by the PTF project. The observation numbers in the density map of sky
covered by PTF/iPTF are shown in Figure~\ref{fig:ptf_field}.

\subsection{Catalina Real-Time Transient Survey}
The CRTS project is conducted by analyzing data from the Catalina Sky Survey
(CSS), which is originally designed for the study of asteroids. The CRTS team
made use of the photometric data for studying the transient sky
\citep{2009ApJ...696..870D, 2011BASI...39..387M,
2012MAXIConf}.\footnote{\url{http://crts.caltech.edu}}  The data was gathered
using three telescopes in Northern and Southern Hemispheres, including the
Catalina Sky Survey (CSS, 0.7m), the Mt. Lemmon Survey (MLS, 1.5m), and the
Sliding Springs Survey (SSS, 0.5m).  No filter was adopted for the survey to
maximize the discovery of asteroids.  The CRTS data is available to the public
through the Catalina Surveys Data Release 2 (CSDR2)
website.\footnote{\url{http://nesssi.cacr.caltech.edu/DataRelease/}} The
calibrated light curves can be accessed by users through the interface.

\subsection{Lulin One-Meter Telescope}
The orbital periods of the CVs are essential to the discussion on the mechanism
of their long-term periodicities.  For our sources of interest, only a portion
of them have known orbital periods, as presented in \citet{2006yCat.5123....0D}
and references therein.  To investigate the targets with unknown orbital
periods, we used the Lulin One-Meter Telescope, located in central Taiwan, to
perform short-cadence observations.  The LOT is a telescope with a
field-of-view (FOV) of $11 ^\prime\times 11 ^\prime$. The limiting magnitudes
with 5 min exposures are approximately $20 - 20.5$ mag in V and R bands.  We
used the R-band filter for our short-cadence observations.  The 3 sources we
observed with LOT are presented in Table~\ref{tab:optlog}.

\begin{table}[h!]
\centering
\caption{Observation log of our LOT observations} \label{tab:optlog}
\begin{tabular}{crcccc}
    \decimals
\hline
\hline
    Source & Nights & Cadence & Exposure & Filter & $N$ \\
    & (day) & (second) & (second) & &  \\
\tableline
    UMa 01 & 5 & 210 & 200 & R & 403 \\
    CT Boo & 4 & 210 & 200 & R & 140 \\
    Her 12 & 2 & 310 & 300 & R & 105 \\
\tableline
\multicolumn{6}{p{0.6\textwidth}}{ Note: Cadence is the planned observation
interval between each of the each frames, and is possibly affected by weather
conditions or the instrument. $N$ is the number of the observations.}
\tablewidth{0pt}
\end{tabular}
\end{table}

\section{Data Analysis} \label{sec:analy}
\subsection{Photometric Calibration} \label{sec:calib}
The photometric systems of PTF and CRTS are different.  The light curves from
the PTF repository are in the Mould R band.  The data retrieved from the CSDR2
are in white light (unfiltered).  To calibrate the difference between the
systems, a linear relation between the PTF and CRTS systems was assumed in our
analysis.  The linear regression to find the cofficients of the line is
described as: $M_\mathrm{PTF} = A \times M_\mathrm{CRTS} + B$, where
$M_\mathrm{PTF}$ is the $R$ band magnitude measured by the PTF, and
$M_\mathrm{CRTS}$ is the white-light magnitude measured by the CRTS.  The slope
$A$ and the intercept $B$ were fit across the magnitudes of different reference
stars in the nearby regions of the CVs.  Most of the resulting fit parameters
$A$ and $B$ are consistent across the entire data set.  The magnitudes of the
CVs were shifted to the PTF photometric system by the linear relation with the
best-fit parameters.  The photometric results from the relative photometry are
sufficient for further temporal analysis.  Therefore, global calibration on the
photometric systems is not necessary.

\subsection{Timing Analysis} \label{sec:timing}
\begin{figure}[ht!]
\plotone{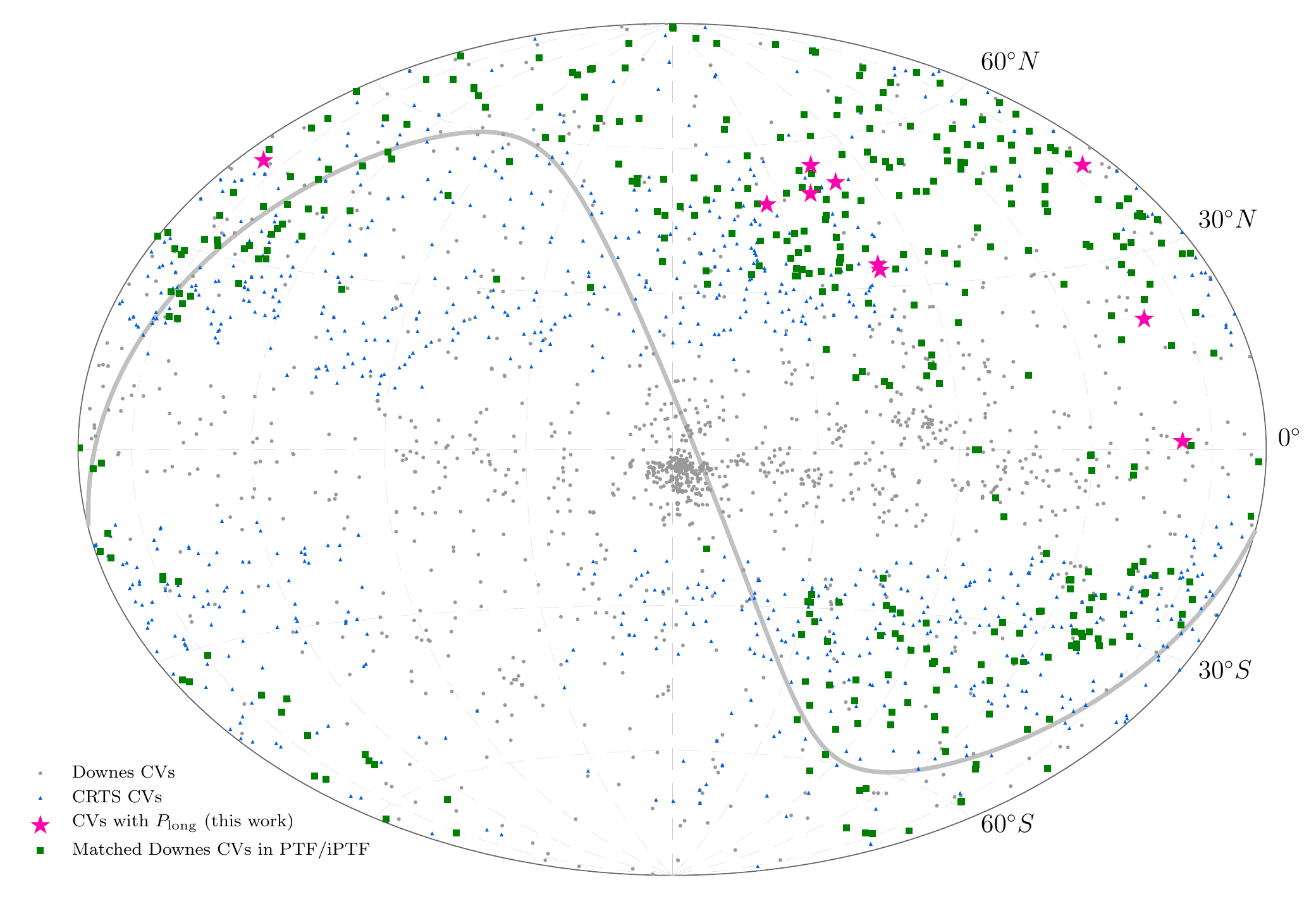}
\caption{Spatial distribution of Galactic CVs found by various previous studies
    and in this study.  Green squares represent the matched sources in PTF with
    the catalog from \citet{2006yCat.5123....0D}.  Gray filled circles are
    targets in \citet{2006yCat.5123....0D} catalog but not matched in the PTF
    survey. Blue triangles are the CV candidates with the CRTS project,
    proposed by \citet{2014MNRAS.441.1186D}. The ecliptic plane is drawn with a
    gray line.  This figure is presented in Galactic coordinates with the Hammer
    projection.  \label{fig:cv_match}}
\end{figure}

We regarded the data from PTF as the major data set, and the data from CRTS as
supplemental.  We cross-matched the Downes' catalog and the PTF database, and
found that 344 CVs have PTF photometric data. The spatial distribution of the
matches, along with all the CVs in the Downes' catalog, are shown in
Figure~\ref{fig:cv_match}.  About 22\% of the CVs in Downes' catalog have a PTF
counterpart. This is because the photometric catalogs of the PTF contain mostly
fields in the non-Galactic plane, owing to the difficulties in processing
Galactic-plane data and the restrictions inherent to the aperture photometric
pipeline. 

\begin{figure*}
\gridline{
    \fig{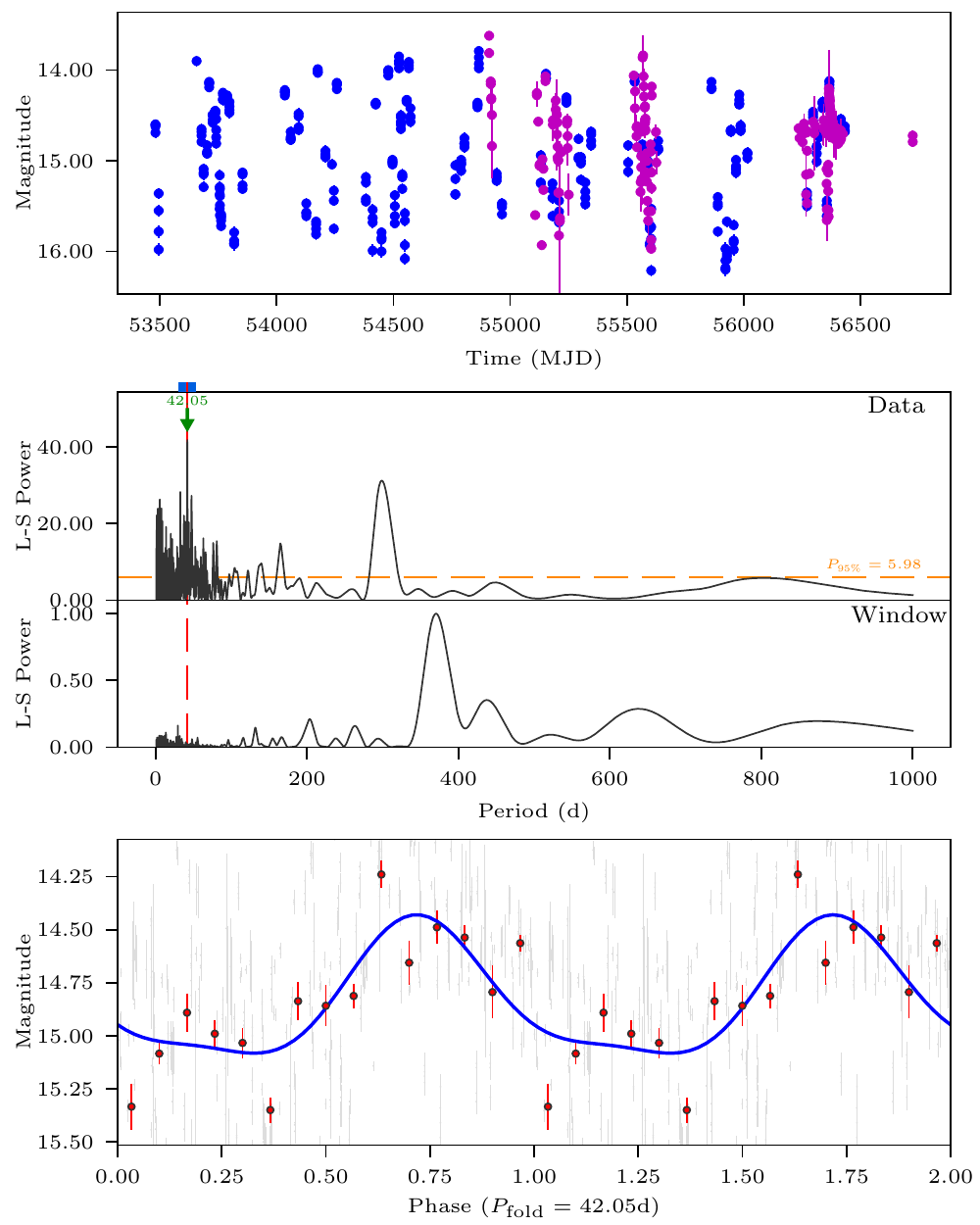}{0.4\textwidth}{\scriptsize (a) BK Lyn ($P_\mathrm{long} = 42.05 \pm
    0.01 ~\mathrm{d}$)}
    \fig{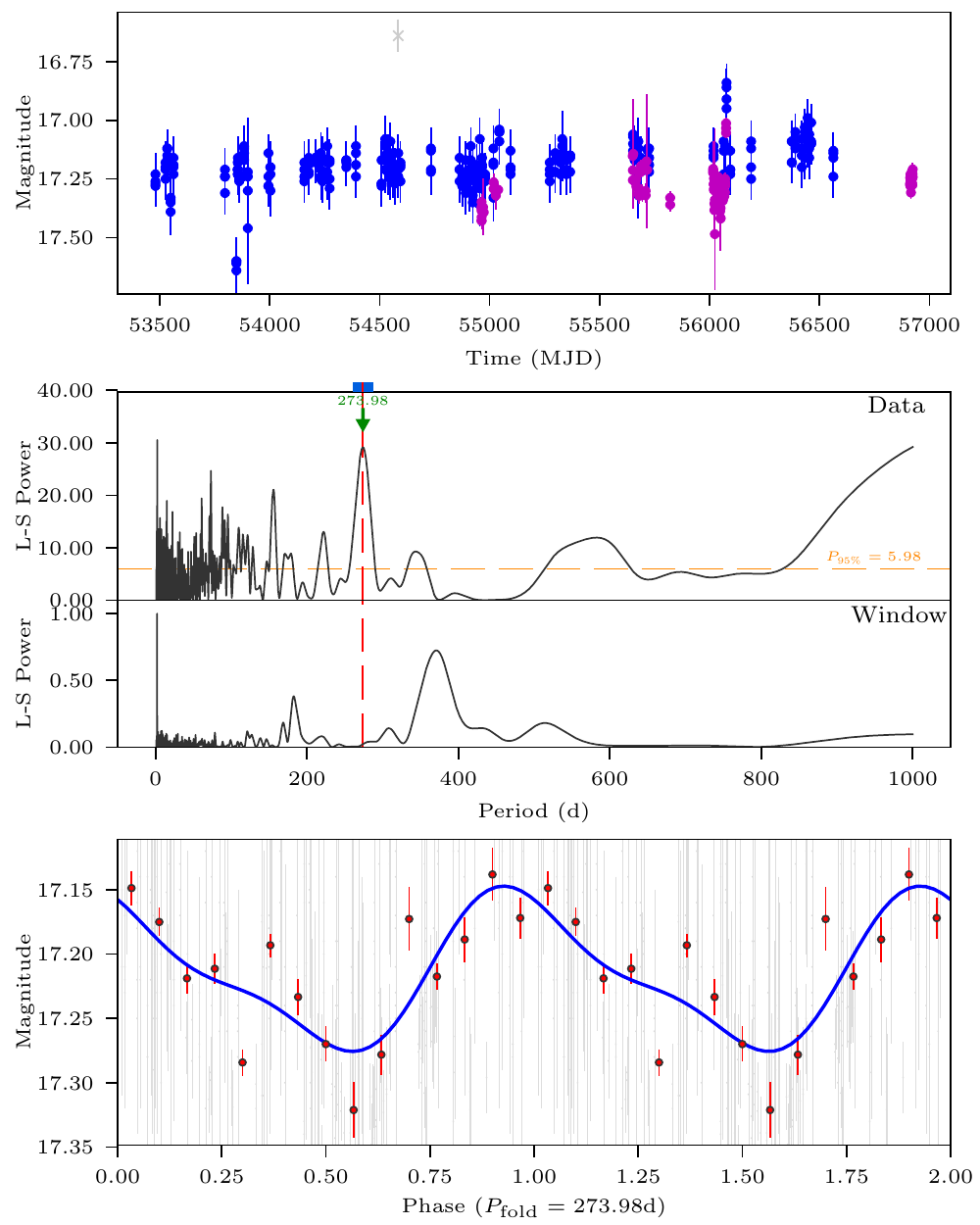}{0.4\textwidth}{\scriptsize (b) CrB 06 ($P_\mathrm{long} = 273.98\pm 
    2.17~\mathrm{d}$)}
}
\gridline{
    \fig{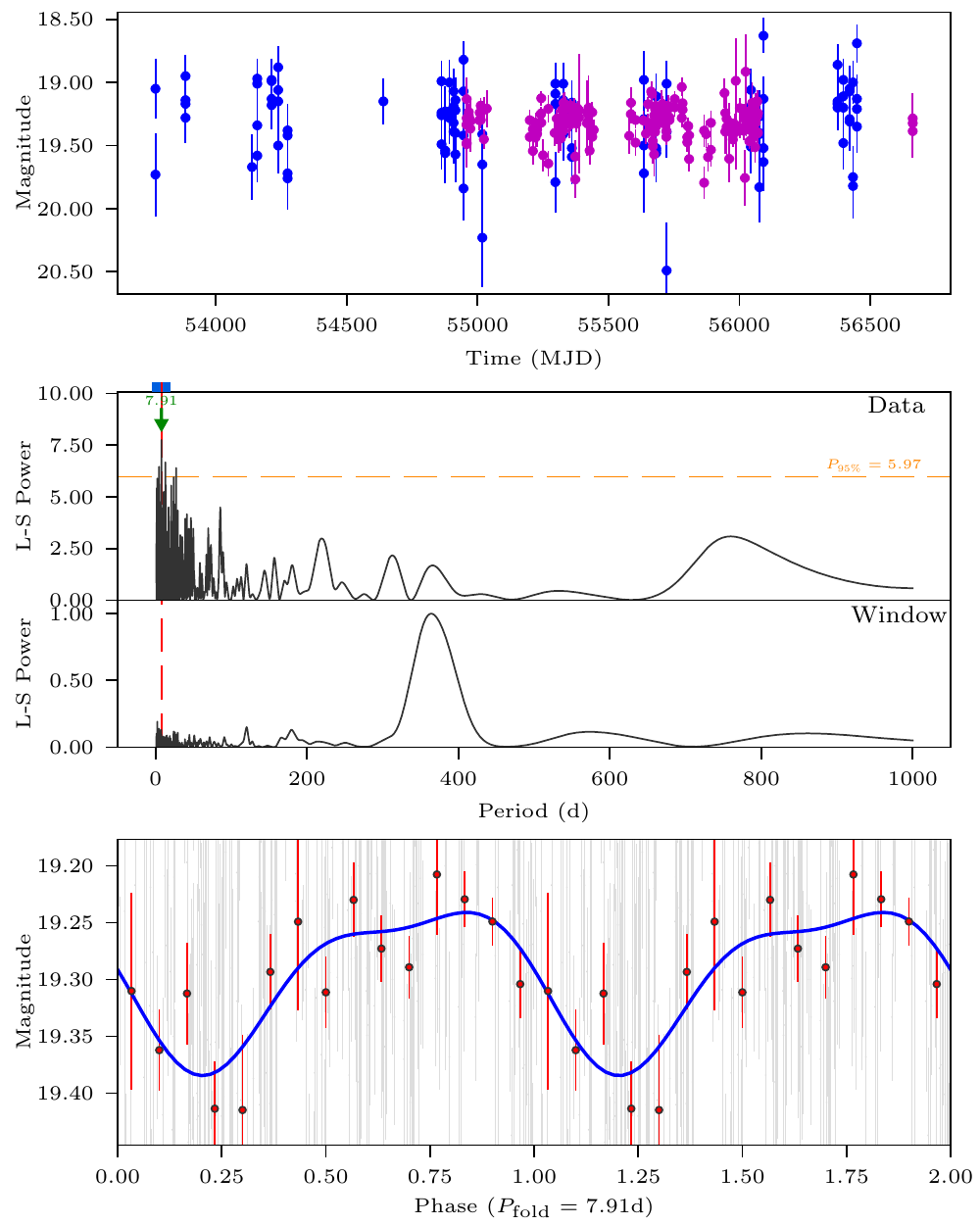}{0.4\textwidth}{\scriptsize (c) CT Boo ($P_\mathrm{long} = 7.91\pm
    1.05~\mathrm{d}$)}
    \fig{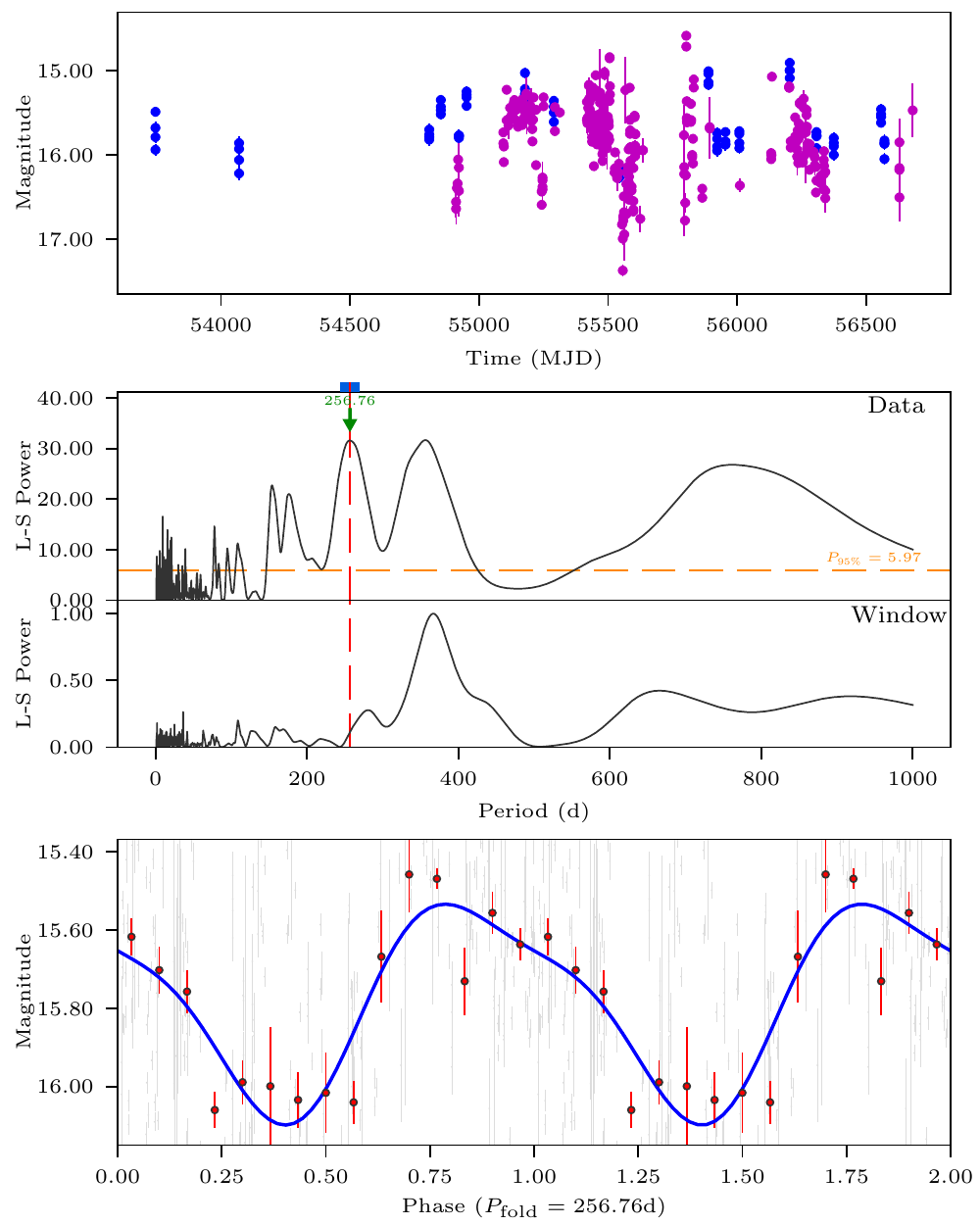}{0.4\textwidth}{\scriptsize (d) LU Cam ($P_\mathrm{long} = 256.76\pm
    2.10 ~\mathrm{d}$)}
}
\caption{\scriptsize Upper panel -- light curves from PTF and CRTS.  Magenta:
    PTF; blue: CRTS; gray: data points of outbursts or dips eliminated in the
    analysis. Middle panel -- Lomb-Scargle periodogram of the joint light curve
    (upper), and the window function (lower, normalized). Blue bar in the top
    of the panel shows the error from the Monte Carlo simulation. Green arrows
    show the values of the periods of our interests. Orange dashed line
        indicates the 95\% confidence level ($P_{95\%}$). The red dashed
        vertical line indicates the period used for the folded light curve in
        the lower panel. Lower panel -- gray: folded light curves; red: binned
        light curves with 15 bins to address the modulation; blue: the fit line
        with two sinusoidal components.  \label{fig:flc1}}
\end{figure*}

\begin{figure*}
\gridline{
    \fig{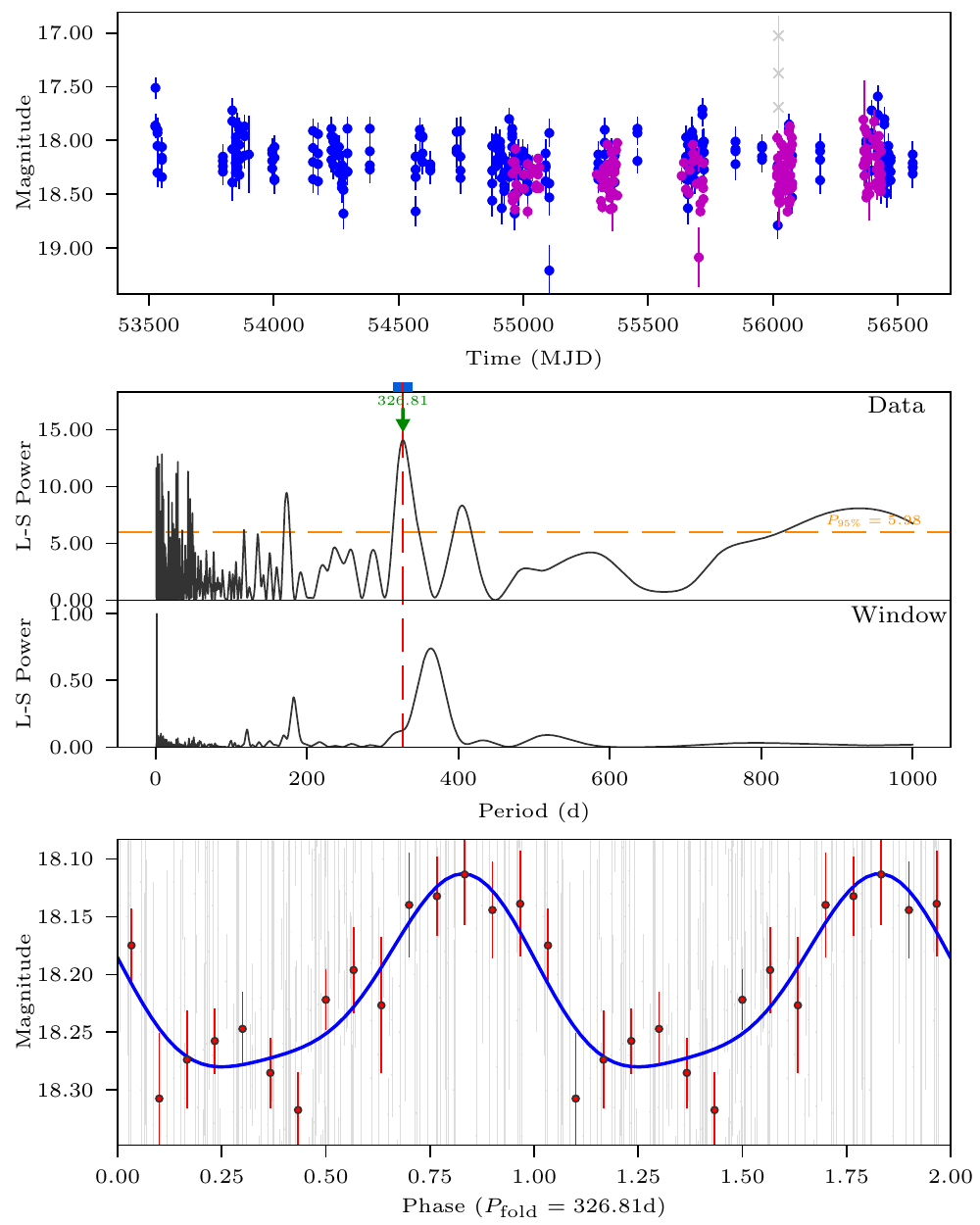}{0.4\textwidth}{\scriptsize (e) Her 12 ($P_\mathrm{long} = 326.81
    \pm 1.32 ~\mathrm{d}$)} 
    \fig{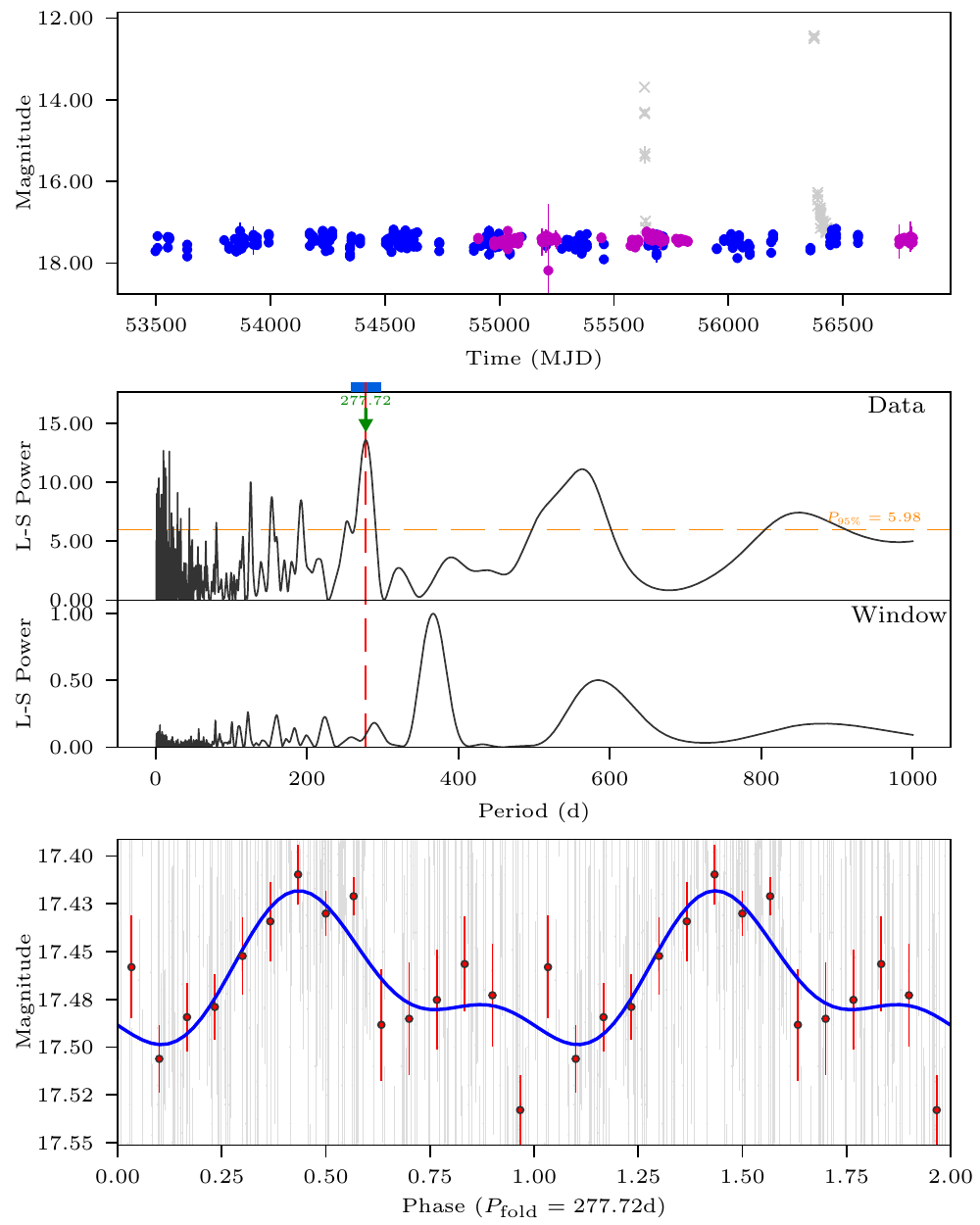}{0.4\textwidth}{\scriptsize (f) QZ Ser ($P_\mathrm{long} =
    277.72 \pm 8.76~\mathrm{d}$)}
}

\gridline{
    \fig{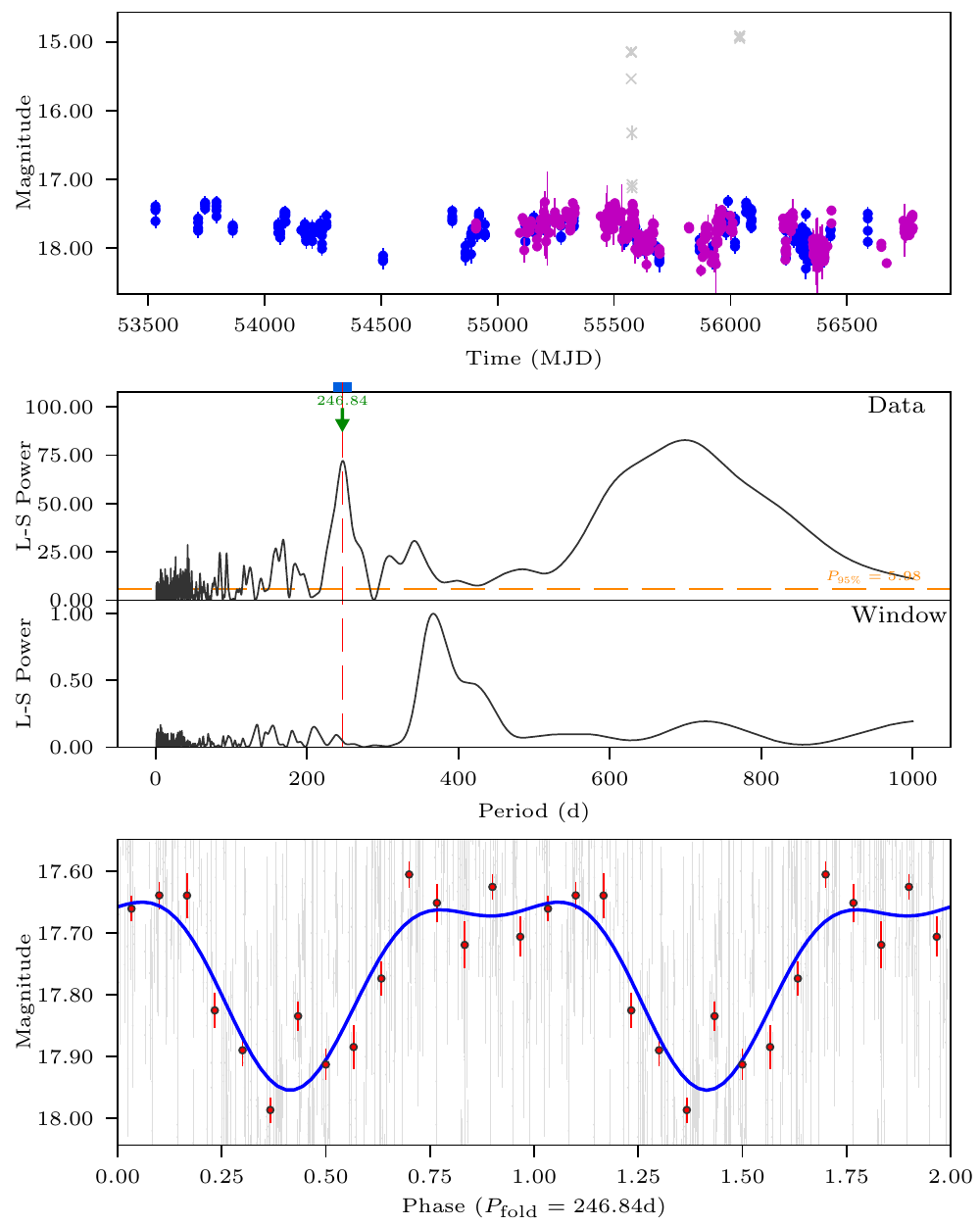}{0.4\textwidth}{\scriptsize (g) UMa 01 ($P_\mathrm{long} =
    246.84 \pm 0.81 ~\mathrm{d}$)}
    \fig{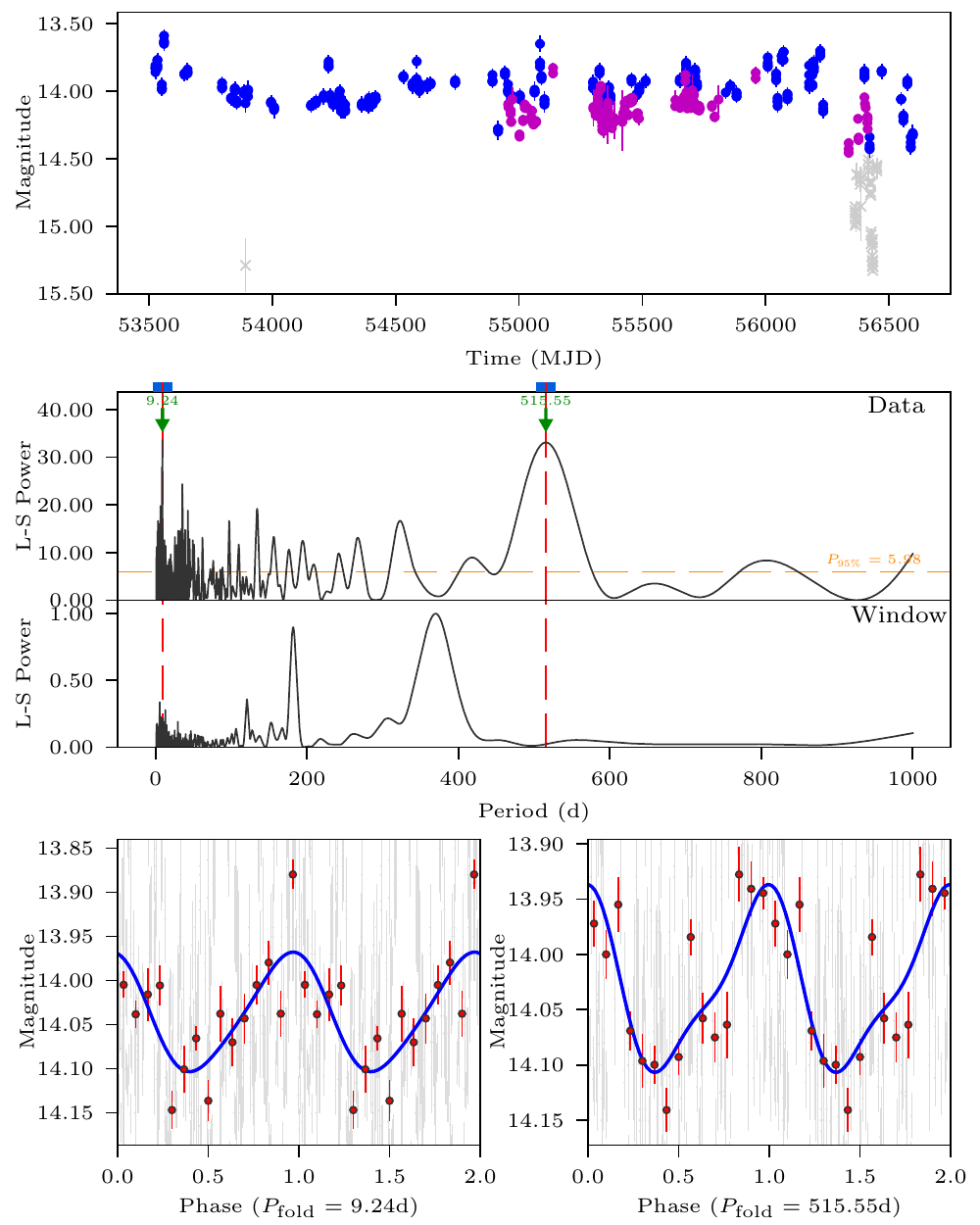}{0.4\textwidth}{\scriptsize (h) V825 Her ($P_\mathrm{long} =
    515.55 \pm 1.85 ~\mathrm{d}$ or $P_\mathrm{long} =
    9.24 \pm 0.05 ~\mathrm{d}$)}
}
\caption{Light curves (upper panel), Lomb-Scargle periodogram of CVs (upper
    of middle panel) and window function (lower of middle panel), and folded
    light curves (lower panel) (continued). The descriptions and labels are the
    same as in Figure~\ref{fig:flc1}.  \label{fig:flc2}}
\end{figure*}

\begin{figure*}
\gridline{
    \fig{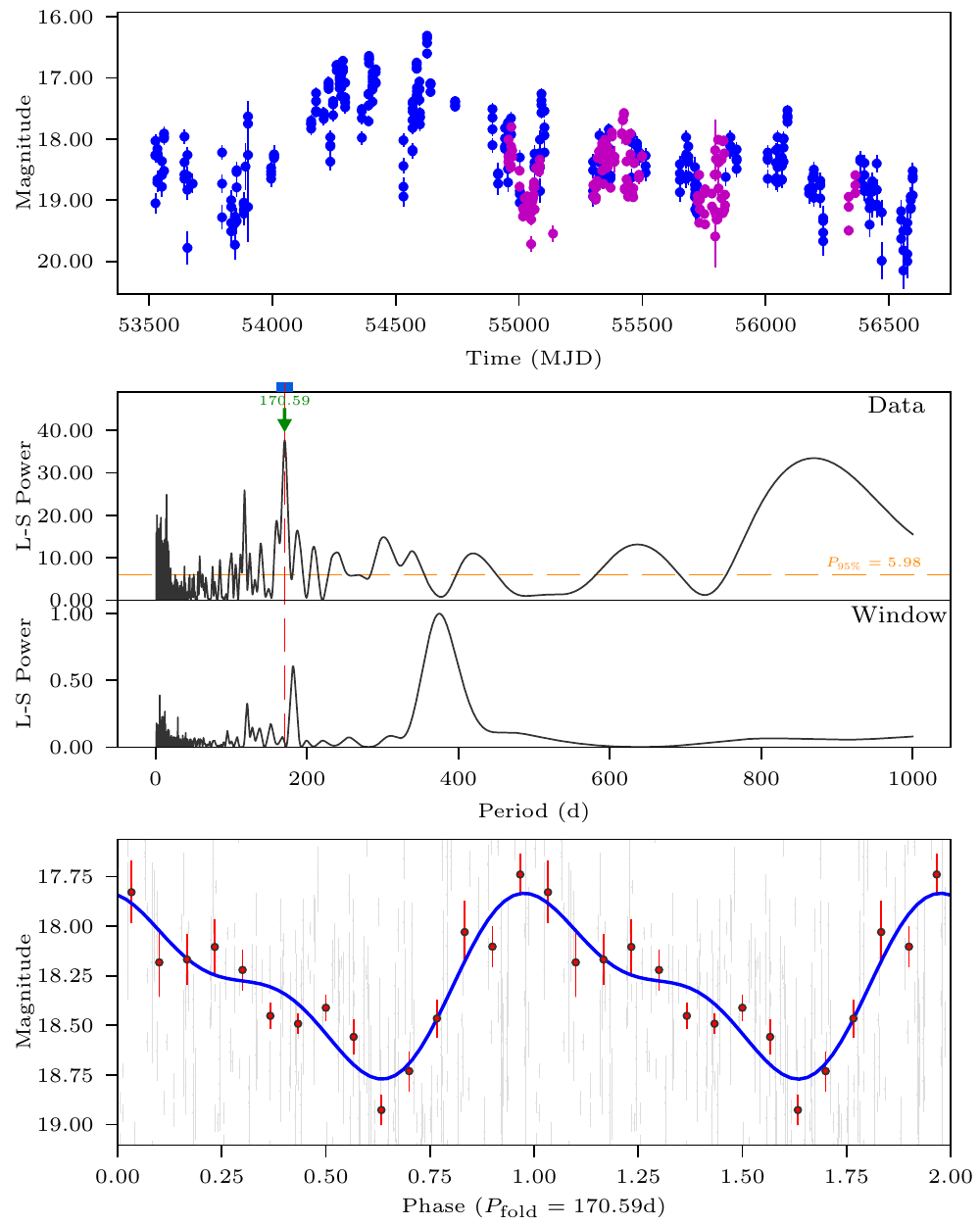}{0.4\textwidth}{\scriptsize (i) V1007 Her ($P_\mathrm{long} =
    170.59 \pm 0.12~\mathrm{d}$)}
    \fig{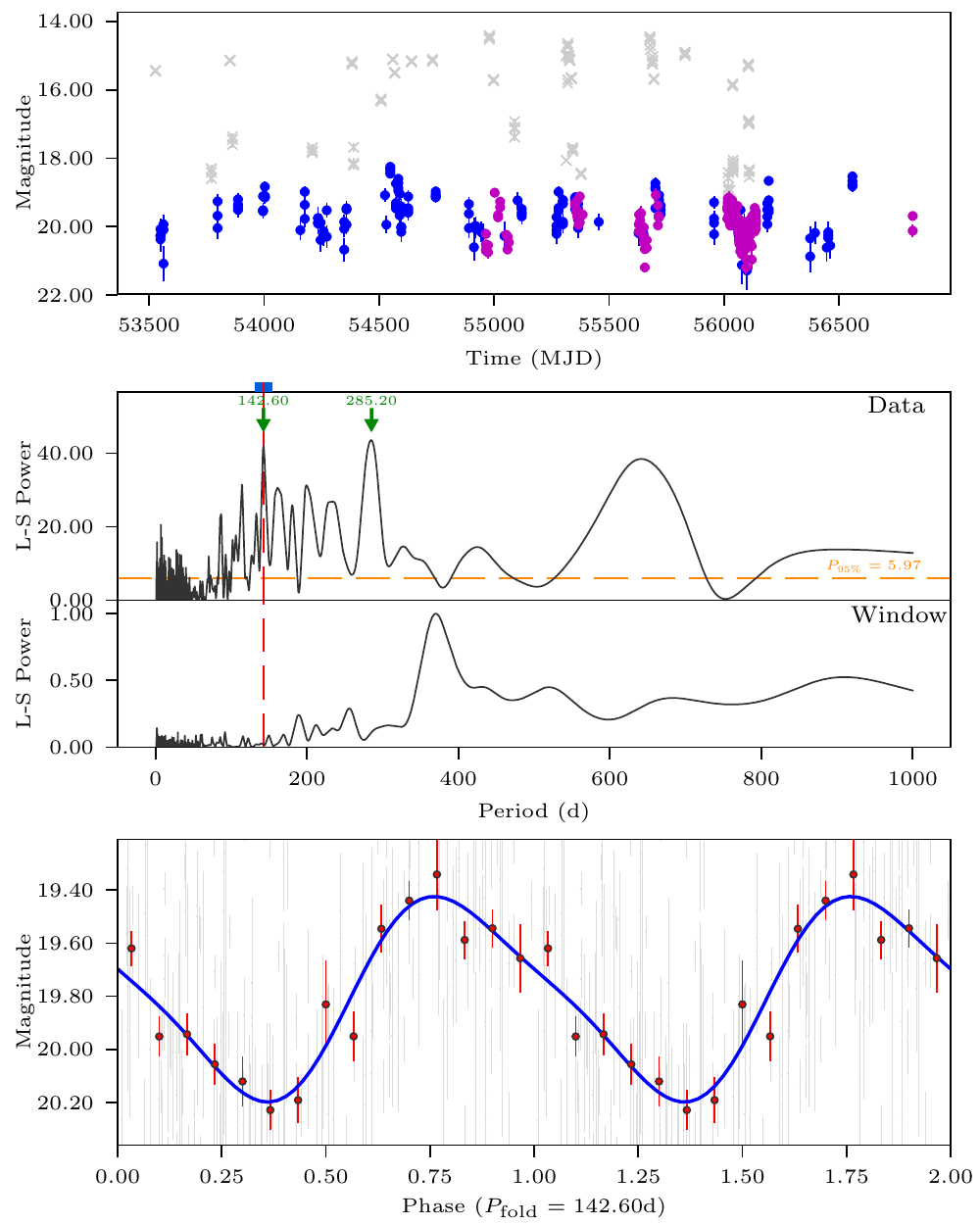}{0.4\textwidth}{\scriptsize (j) VW CrB ($P_\mathrm{long} =
    142.60 \pm 0.24~\mathrm{d}$)}
}
\caption{Light curves (upper panel), Lomb-Scargle periodogram of CVs (upper
    of middle panel) and window function (lower of middle panel), and folded
    light curves (lower panel) (continued). The descriptions and labels are the
    same as in Figure~\ref{fig:flc1}.  \label{fig:flc3}}
\end{figure*}

Matched CVs with only a few observations are not sufficient for studying their
timing properties. Therefore, we selected the light curves with more than 100
observations for further investigation.  About 100 of the matched CVs satisfied
this criterion.  These light curves were combined with their corresponding
light curves in CRTS.  In addition, some of the sources exhibit outbursts in
the joint light curves. To avoid the interference from the outbursts, the data
points for the outbursts in all the joint light curves were eliminated by
visual inspection before the analysis.  The dip features that are present in
the light curves of cataclysmic variable QZ Ser were eliminated as well,
because they are aperiodic.  The joint light curves of the 344 selected CVs
were then adopted for periodicity analysis.  There are 10 CVs found to have
long-term periodicities.  The joint light curves of these 10 CVs are presented
in the upper panels of Figure~\ref{fig:flc1}, \ref{fig:flc2}, and
\ref{fig:flc3}.  The total time span, the data points for timing analysis, and
the outburst numbers for the light curves are listed in
Table~\ref{tab:summary}. 

\begin{longrotatetable}
\tabletypesize{\scriptsize}
\begin{deluxetable*}{ccccccrrrrrrrc}
\tablecaption{Summary of CVs having long-term periodicities \label{tab:summary}}
\tablehead{
    \colhead{CV Name\tablenotemark{a}}  &
    \colhead{RA (J2000)} &
    \colhead{Dec (J2000)} & 
    \colhead{Alternative Name\tablenotemark{b}} & 
    \colhead{Type\tablenotemark{c}}  &
    \colhead{$P_\mathrm{orb}$\tablenotemark{d}} & 
    \colhead{$\sigma_{P_\mathrm{orb}}$\tablenotemark{e}} & 
    \colhead{$P_\mathrm{long}$\tablenotemark{f}} &
    \colhead{$\sigma_{P_\mathrm{long}}$\tablenotemark{g}} & 
    \colhead{$p({P_\mathrm{long}})$\tablenotemark{h}} & 
    \colhead{Amp.\tablenotemark{i}} &
    \colhead{T$_\mathrm{span}$\tablenotemark{j}} &
    \colhead{N\tablenotemark{k}} &
    \colhead{$N_\mathrm{ob}$ \tablenotemark{l}} 
    \\
    \colhead{} & \colhead{(hh:mm:ss)} & \colhead{($\pm$dd:mm:ss)} & \colhead{} &
    \colhead{} & \colhead{(min)} & \colhead{(min)} & \colhead{(day)} &
    \colhead{(day)} & \colhead{} & \colhead{(mag)} & \colhead{(day)} & \colhead{(\#)} &
    \colhead{(\#)}
}
\startdata
BK Lyn    &  09:20:11.20 & +33:56:42.3  & 2MASS J09201119+3356423  & DN/NL & 
107.97\tablenotemark{m}   & 0.07\tablenotemark{m} & \textbf{42.05}  & \textbf{0.01} & \textbf{1.88e-20} &
\textbf{0.65} & \textbf{3242.92} & \textbf{514} & --- \\
CrB 06    &  15:32:13.68 & +37:01:04.9 & 2MASS J15321369+3701046 & NL    & 
---  & --- &  \textbf{273.98}  & \textbf{2.17} & \textbf{1.18e-13} &
\textbf{0.11} & \textbf{3443.86} & \textbf{400} & \textbf{1} \\
CT Boo    &  14:08:20.91 & +53:30:40.2  & --- & NL & 
\textbf{230.72} & \textbf{1.78} & \textbf{7.91} & \textbf{1.05} &
\textbf{4.11e-04} &
\textbf{0.13} & \textbf{2893.00} & \textbf{296} & --- \\
  &  & &&& 
\textbf{78.65} & \textbf{0.36} & & & &
&&&\\
LU Cam    &  05:58:17.89 & +67:53:46.0  & 2MASS J05581789+6753459  & DN     & 
215.95\tablenotemark{n}  & 0.0005\tablenotemark{n} & \textbf{256.76} & \textbf{2.10} & \textbf{3.18e-03} &
\textbf{0.62} & \textbf{2934.95} & \textbf{363} & --- \\
Her 12    &  15:50:37.28 & +40:54:40.0  & SDSS J155037.27+405440.0 & CV     & 
\textbf{75.62}      & \textbf{0.008} & \textbf{326.81} & \textbf{1.32} &
\textbf{4.97e-07} &
\textbf{0.18} & \textbf{3033.88} & \textbf{429} &  \textbf{1} \\
&&&&& 
\textbf{173.65}      & \textbf{1.57} &&&&
&&&\\
QZ Ser    &  15:56:54.47 & +21:07:19.0 & SDSS J155654.47+210719.0 & DN     & 
119.75\tablenotemark{o}  & 0.002\tablenotemark{o} & \textbf{277.72} & \textbf{8.76} & \textbf{9.18e-07} &
\textbf{0.09} & \textbf{3308.01} & \textbf{510} & \textbf{2} \\
UMa 01    &  09:19:35.70 & +50:28:26.2 & 2MASS J09193569+5028261  & CV     &
\textbf{404.10}    & \textbf{0.30} & \textbf{246.84}  & \textbf{0.81} &
\textbf{8.01e-20} &
\textbf{0.44} & \textbf{3251.05} & \textbf{601} & \textbf{1} \\
V825 Her  &  17:18:36.99 & +41:15:51.2 & 2MASS J17183699+4115511 & NL     &
296.64\tablenotemark{p}  & 2.88\tablenotemark{p} & \textbf{515.55} &
\textbf{1.85} & \textbf{3.28e-16} &
\textbf{0.16} & \textbf{3069.71} & \textbf{510} &  --- \\
&&&&&
& & \textbf{9.24} & \textbf{0.05} & \textbf{4.51e-16} &
\textbf{0.18} & && \\
V1007 Her &  17:24:06.32 & +41:14:10.1 & 1RXS J172405.7+411402 & AM &
119.93\tablenotemark{q}  & 0.0001\tablenotemark{q} & \textbf{170.59} & \textbf{0.12} & \textbf{2.13e-18} &
\textbf{0.90} & \textbf{3069.71} & \textbf{507} & --- \\
VW CrB    &  16:00:03.71 & +33:11:13.9 & USNO-B1.0 1231-00276740 & DN     &
---      & --- & \textbf{142.60} & \textbf{0.24} & \textbf{1.92e-21} &
\textbf{0.87} & \textbf{3271.19} & \textbf{351} & \textbf{$>$10} \\
\enddata
 \tablecomments{The numbers in bold face indicate the values derived in this work.
 Multiple possible orbital periods present in CT Boo and Her 12, and they are
 listed in the table accordingly. \\\\
 $^a$ CV name: name designated in this project; 
 $^b$ Alternative name: the source name in 2MASS, SDSS, ROSAT
 bright source catalog (1RXS) or USNO B1.0 catalogs;
 $^c$ Type: the type of CVs designated in
     \citet{2006yCat.5123....0D}(NL: nova-like, DN: dwarf nova, AM: AM Her
 (polar));
 $^d$ $P_\mathrm{orb}$: orbital periods;
 $^e$ $\sigma_{P_\mathrm{orb}}$: errors of orbital periods; 
 $^f$ $P_\mathrm{long}$: long-term periods derived in this work;
 $^g$ $\sigma_{P_\mathrm{long}}$: errors of long-term periods derived in this work;
 $^h$ $p(P_\mathrm{long})$: the $p$-value of the long-term period;
 $^i$ Amplitude: peak-to-peak magnitude difference from the fit light curve;
 $^j$ $T_\mathrm{span}$: total time span;
 $^k$ $N$: number of observations excluding outbursts and dips;
 $^l$ $N_\mathrm{ob}$: number of outbursts observed with PTF and CRTS;
 $^m$ \citet{1996MNRAS.278..125R};
 $^n$ \citet{2007PASP..119..494S};
 $^o$ \citet{2002PASP..114.1117T};
 $^p$ \citet{2003AAS...203.4407R};
 $^q$ \citet{1998MNRAS.296..437G}}
\end{deluxetable*}
\end{longrotatetable}

To study the timing properties of these CVs, the Lomb-Scargle periodogram
\citep[LS,][]{1976Ap&SS..39..447L,1982ApJ...263..835S} was adopted to search
for the periodicities, and the method of Phase-Dispersion Minimization
\citep[PDM,][]{1978ApJ...224..953S} was used to cross-check the candidate
periods.  The periods obtained from these two methods are consistent to each
other.  In this paper, we only present the results from LS.  For the same FOV,
the observation schedules of PTF and CRTS are not regular and not fixed.
Therefore, there are many observation gaps in the light curves.  Aliases can be
observed in the power spectra owing to the gaps in the observations, and this
makes analyzing the light curves difficult.  To distinguish and eliminate the
aliasing introduced by the windows, the periodogram of the observation windows
are plotted against the source periodogram, as demonstrated in the lower graphs
of the middle panels in Figure~\ref{fig:flc1}, \ref{fig:flc2}, and
\ref{fig:flc3}.

A summary of the possible CVs with the long-term periodicities are listed
in Table~\ref{tab:summary}.  For the LS periodogram, the $p$-value is
frequently adopted to represent the detection significance of a periodic
signal. The $p$-value is defined as the probability that the periodogram power
can exceed a given value $P_i$, expressed as $\mathrm{Prob}(P > P_i)$, and can
be expressed as:
\begin{equation} 
    p \equiv \mathrm{Prob}(P > P_i) = \left( 1 - \frac{2P_i}{N-1} \right)^{\frac{N-3}{2}}
    \label{eq:pvalue}
\end{equation}
where $N$ is the total number of observations \citep{2009A&A...496..577Z}.

To derive the corresponding uncertainty of the most probable long-term period,
Monte Carlo simulations were conducted $10^4$ times under the assumption that
the observational errors are Gaussian distributed.  On the basis of the
observed magnitudes and corresponding errors of each CV, we generated $10^4$
simulated light curves.  The root-mean-square values of the peak values in the
power spectra were adopted as the statistic errors of the derived periods. The
errors of the periods are also listed in Table~\ref{tab:summary}. 

Light curves were folded with the most probable periods in the power spectra,
as shown in the lower panels of Figure~\ref{fig:flc1}, \ref{fig:flc2}, and
\ref{fig:flc3}.  To show the profiles of the long-term modulations, we binned
the folded light curves into 15 phase intervals, and fit each of them with a
two-sinusoidal function, which is given as:
\begin{equation}
    M = c_1 \sin(2\pi \phi + c_2) + c_3 \sin(4 \pi \phi + c_4) + c_5
\end{equation}
where $\phi$ is the phase of the folded light curve, $M$ is the
corresponding magnitude, and $c_1$ through $c_5$ are the fit cofficients.

The scattering of the points in the folded light curves may be caused by the
original photometric uncertainties. However, the variations from shorter time
scales, such as the orbital modulations, contribute to the light curves;
therefore, these variations cannot be neglected.  We tried to fold the original
light curves of the CVs with their known constant orbital periods; however,
this failed in reconstructing their orbital profiles.

\subsection{Lulin One-Meter Telescope Observations} \label{sec:lot}
\begin{figure*}
\gridline{
    \fig{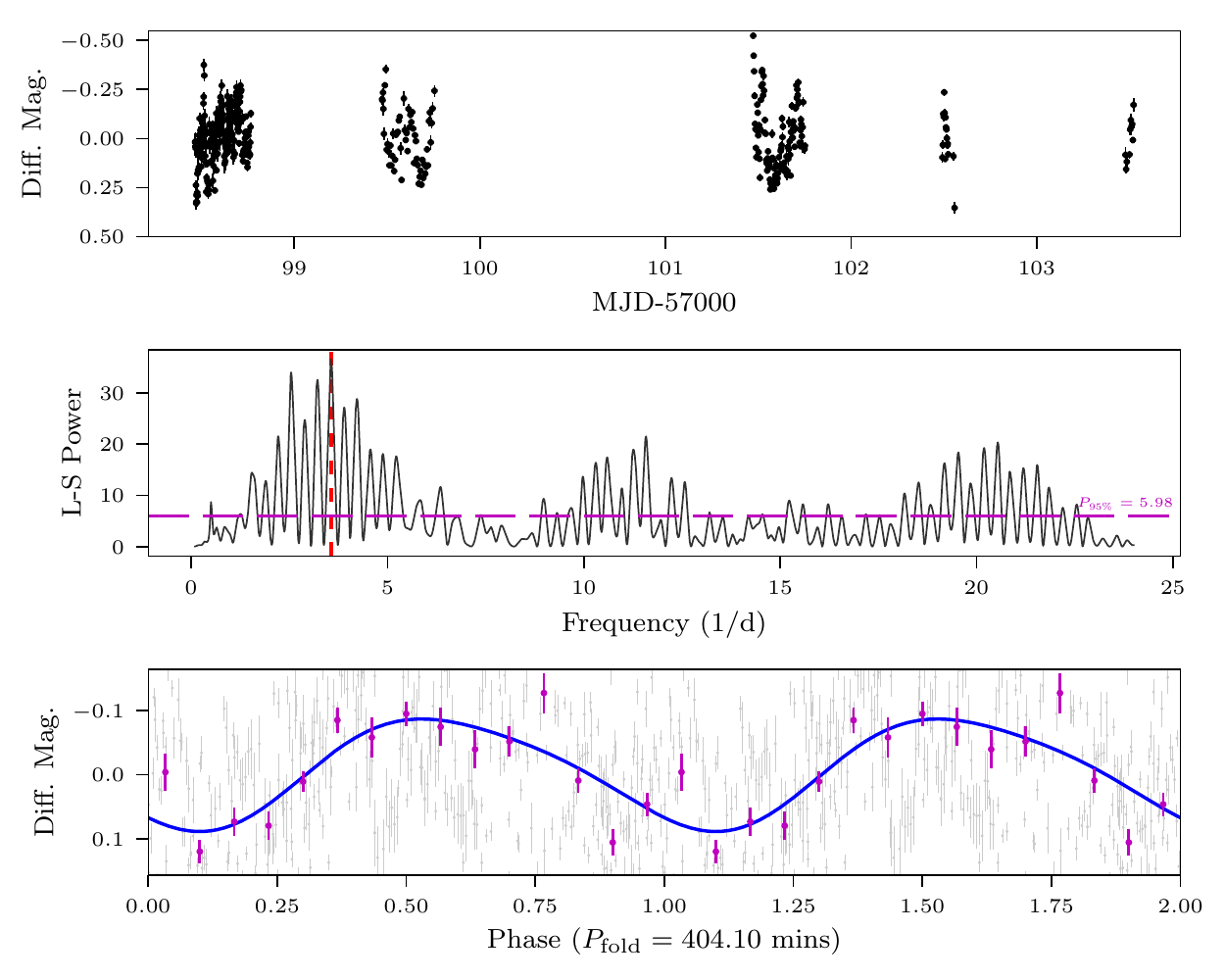}{0.5\textwidth}{(a) UMa 01 ($P_\mathrm{orb} =
    404.10\pm0.30~\mathrm{min}$)}
    \fig{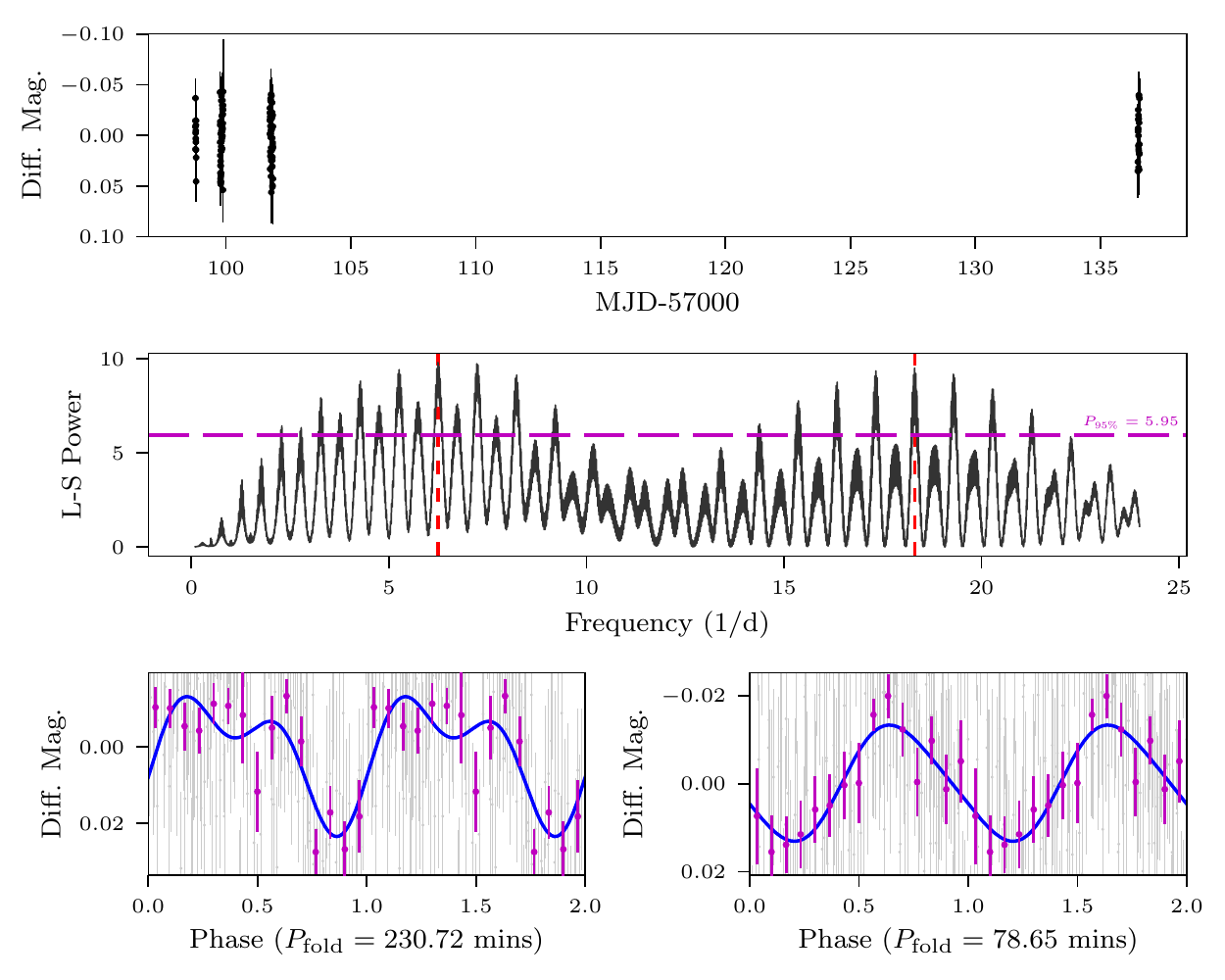}{0.5\textwidth}{(b) CT Boo ($P_\mathrm{orb} =
    230.72\pm1.78$ or $78.65\pm0.36~\mathrm{min}$)}
}
\gridline{
    \fig{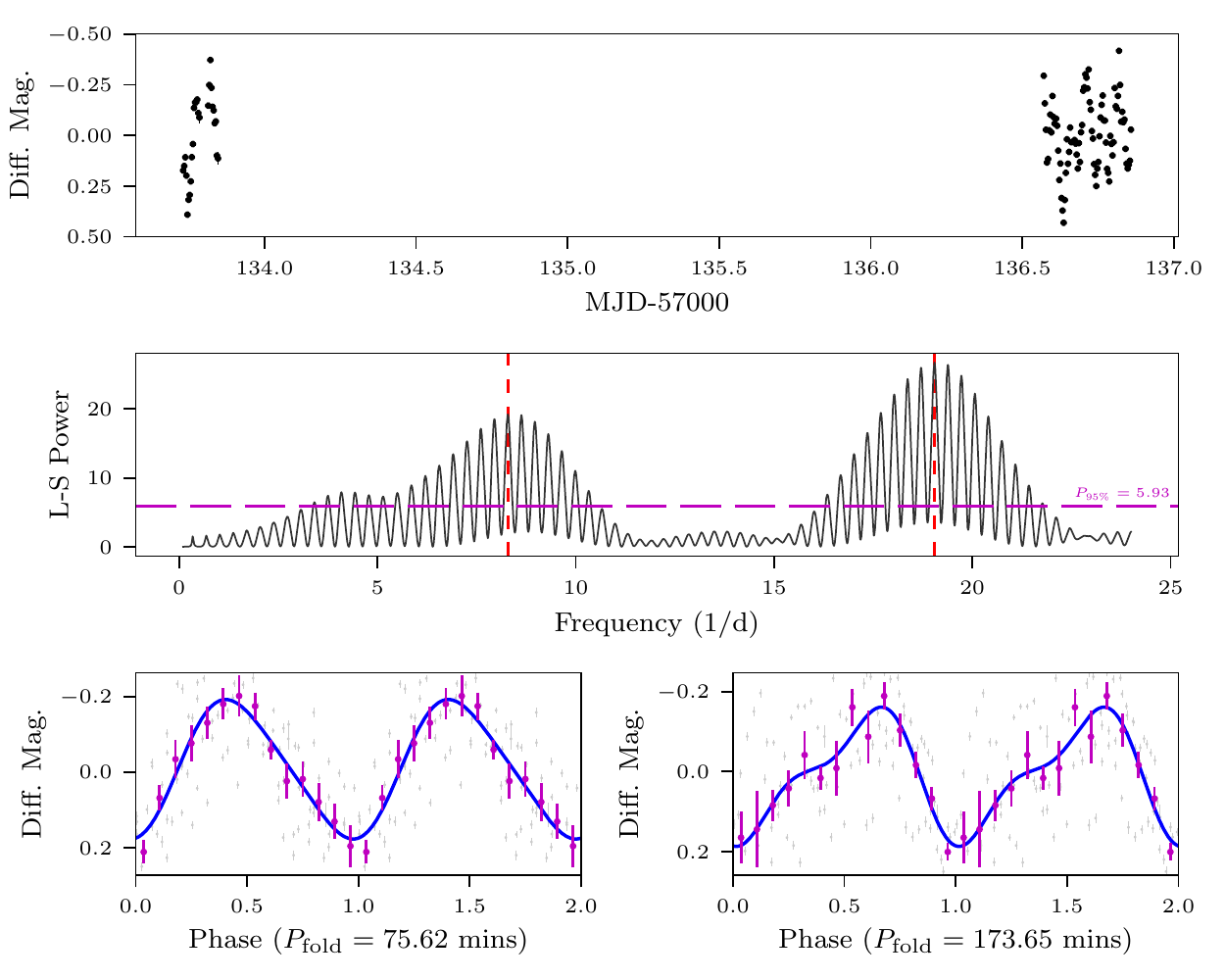}{0.5\textwidth}{(b) Her 12 ($P_\mathrm{orb} =
    75.62\pm0.008$ or $173.65\pm1.57~\mathrm{min}$)}
}

\caption{Folded short cadence light curves of (a) UMa 01, (b) CT Boo, and (c)
    Her 12.  Upper panels show the detrended light curves in the available data
    segments.  Middle panels present the Lomb-Scargle periodogram of the CVs.
    The magenta dashed line denotes the L-S power with 95\% confidence level,
    and red dashed lines denote the periodic signals used for the folded
    periods. Lower panels show the folded light curves with the best periods in
    the Lomb-Scargle periodogram. Gray dots are the folded light curves.
    Purple points indicate the binned light curves and errors. Blue curves
    represent the modulation trends with fits to a function with two sinusoidal
    components.  \label{fig:olc}}
\end{figure*}

The images we took using the LOT were reduced by NOAO IRAF
packages.\footnote{\url{http://iraf.noao.edu/}} The images were processed by
subtracting the master bias and dark frames. Then, the flat-field correction
was applied for correcting the pixel-dependent instrument response.  Point
spread function (PSF) fitting was performed on the images using
\texttt{DAOPHOT} to obtain accurate photometric results.  In the light curves
derived from the PSF fitting, different trends exist between different days.
The mean value of the data in each day was subtracted for reducing the
interference of the trend.  The detrended light curves are presented in
Figure~\ref{fig:olc}.  Moreover, the LS periodogram was applied for searching
for periodicities in the light curves.  The folded light curves were produced
by folding the detrended light curves with the best determined periods.  The
power spectra and the folded light curves are shown in Figure~\ref{fig:olc}.
The significance levels are indicated by the $p$-values as well. A summary of
the results is shown in Table~\ref{tab:optres}.

\begin{table}[h!]
\centering
\caption{Periodicities in short-cadence observations} \label{tab:optres}
\begin{tabular}{crrrcc}
\hline
\hline
    Source & $P_\mathrm{opt}$ & $\sigma_{P_\mathrm{opt}}$ & $P_\mathrm{peak}$ & $p$-value & N \\
    & (min) & (min) &&& \\
\tableline
    UMa 01 & 404.10 & 0.30 & 36.54 & 3.74e-18 & 403 \\
\hline
    CT Boo & 230.72 & 1.78 & 9.80 & 2.99e-05 & 140 \\
           & 78.65  & 0.36 & 9.49 & 4.29e-05 & 140 \\
\hline
    Her 12 & 173.65 & 1.57 & 26.75 & 1.00e-16 & 105 \\
           & 75.62 & 0.008 & 19.22 & 6.02e-11 & 105 \\
\tableline
\multicolumn{6}{p{9cm}}{ Note: $P_\mathrm{opt}$: optical periods,
$\sigma_{P_\mathrm{opt}}$: errors in periods by Monte Carlo simulations,
$P_\mathrm{peak}$: peak power in the power spectrum.}
\tablewidth{0pt}
\end{tabular}
\end{table}

Intensive observations were conducted for the unknown orbital periods of the
CVs with long-term periodicities.  In this study, three CVs, namely UMa 01, CT
Boo, and Her 12 were observed and analyzed. UMa 01 (a.k.a. 2MASS
J09193569+5028261) shows a modulation with a period of $404.10\pm0.3$ min.  CT
Boo exhibits two significant periodic signals: $229.13\pm7.02$ and
$79.55\pm0.51$ min. In addition, two significant periods were found in Her 12
(a.k.a. SDSS J155037.27+405440.0): $75.62\pm0.008$ and $173.65\pm1.57$ min.
These periods for the three CVs are all located in the normal orbital period
range of CVs.  However, for the CVs with multiple periodicities, further
investigation is required to distinguish and confirm the orbital periods.
Besides, it is not easy to distinguish them if the modulations in the folded
light curves are orbital humps or superhumps (defined below in
Section~\ref{sec:mec:psup}). We assume that these periods are the possible
orbital periods in the discussion that follows.

\section{Possible Mechanisms of Long-term Periodicity} \label{sec:mechanism}
\begin{figure}[ht!]
\plotone{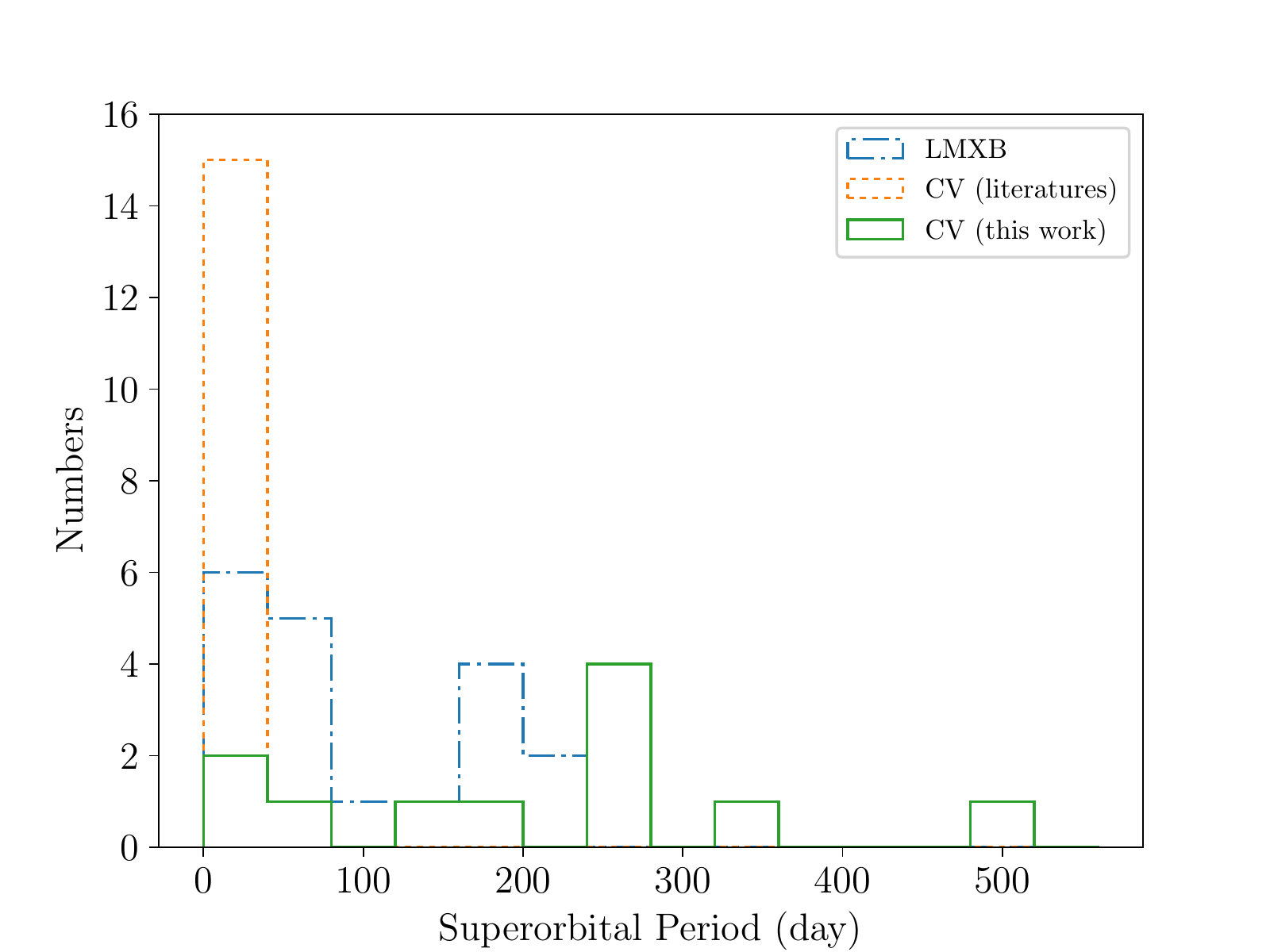}
\caption{Histogram of known superorbital periods in close binary systems.
Blue dashed dots: LMXBs from \citet{2012MNRAS.420.1575K}; orange dots: CVs
from various studies (see Table~\ref{tab:cvprec}); green line: CVs of this
study.} \label{fig:distrib}
\end{figure}

Long-term periodicities are not frequently observed in close binary systems.
However, some X-ray binaries (XBs) exhibit long-term periodicities.
\citet{2012MNRAS.420.1575K} summarized several possible mechanisms for the
long-term variability of XBs, and some of the scenarios may also be applicable
to CV systems.  The distribution of the long-term periodicities in our study is
presented in Figure.~\ref{fig:distrib}. The long-term periods that are
associated with the superhumps of other CVs were gathered from different
literature, and were plotted for comparison.  The long-term periodicities of
the low-mass X-ray binaries (LMXBs) in \citet{2012MNRAS.420.1575K} were
plotted, as well.  In this section, we list and discuss some of the possible
mechanisms that cause the long-term variations in CVs.

\subsection{Precession of Accretion Disk} \label{sec:mec:psup}
CVs are named for their cataclysmic phenomena. The state change of the
accretion disk of CVs may cause their emission flux to change.  In addition to
the orbital modulations, variations in the periods of a few percent longer or
shorter than the orbital periods are called positive and negative superhumps,
respectively.  Superhump was first found in the SU UMa type (a subtype of dwarf
novae (DNe)) of CVs.  During the superoutbursts of a SU UMa type star, a
periodic hump rather than an orbital hump appears and this was called superhump
\citep{Warner1985, 1980A&A....88...66V, 1995CAS....28.....W}. Besides the SU
UMa type of CVs, some of the VY Scl type of CVs (or anti-dwarf novae), a
subtype of nova-like (NL) CV, were detected to have superhump signals. For
example, \citet{2007MNRAS.378..955K} discovered a negative superhump
$P_\mathrm{sh}$ = 3.771 $\pm$ 0.005 h in KR Aur, a VY Scl NL CV. In addition to
the SU UMa and NL systems, superhumps were also observed in the intermediate
polar (IP, or DQ Her type of CVs).  \citet{2012MNRAS.427.1004W} found
superhumps in CC Scl, an IP system, with a period of $P_\mathrm{sh}$ = 1.443 h,
$\sim 4.3\%$ longer than its orbital period $P_\mathrm{orb}$ = 1.383 h.

In general, it is believed that the positive/negative superhump periods are the
beat periods of the orbital period and the precession period of the accretion
disk.  For a positive superhump, the size of the accretion disk is growing as
mass transferring from the companion.  When the radius of the accretion disk is
larger than the 3:1 resonance radius of the binary, the accretion disk becomes
asymmetric and exhibits apsidal precession \citep{1988MNRAS.232...35W,
1989PASJ...41.1005O, 1996PASP..108...39O}.  \citet{1991MNRAS.249...25W}
proposed that this criterion can be achieved only for CV systems with mass
ratio $q \equiv M_\mathrm{d}/M_\mathrm{c} < 0.25 - 0.33$, where $M_\mathrm{c}$
and $M_\mathrm{d}$ are the masses of the accretor and donor, respectively.  A
positive superhump is a consequence of coupling disk apsidal precession and
orbital motion with the beat period for these two types of periodic motions.
On the other hand, a negative superhump is believed to be the result of
coupling between orbital motion and disk nodal precession.

The period excess of a superhump is defined as follows:

\begin{equation}
    \epsilon \equiv \frac{P_\mathrm{sh}-P_\mathrm{orb}}{P_\mathrm{orb}}
    \label{eq:shexcess}
\end{equation}

\begin{figure}[ht!]
\plotone{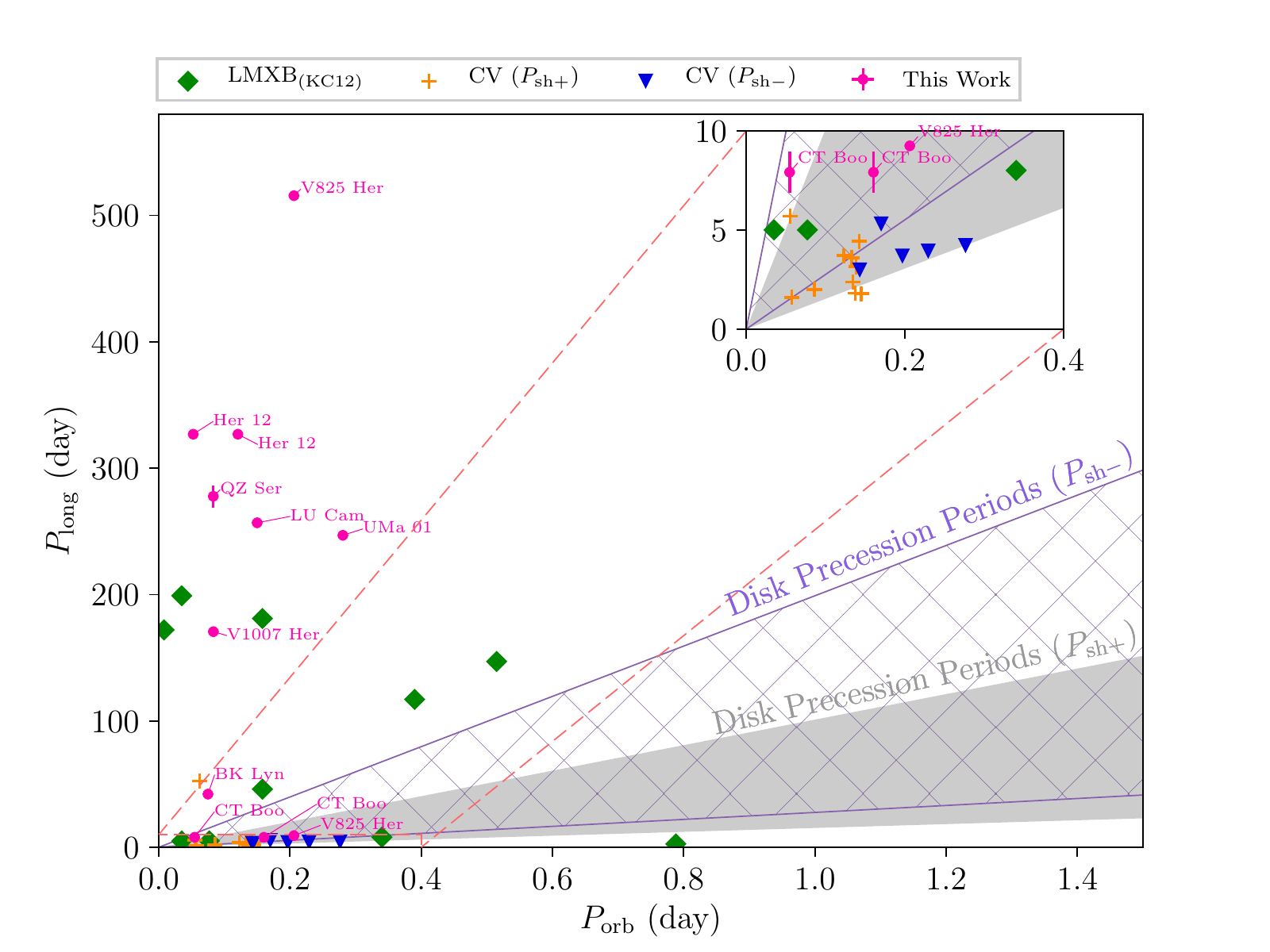}
\caption{$P_\mathrm{long}$ vs. $P_\mathrm{orb}$ relations for LMXBs, and CVs.
LMXBs data are shown with green diamonds (see \citet{2012MNRAS.420.1575K}
(KC12)).  If the long-term periods were provided in ranges, the average values
are adopted for the plot.  CVs from previous studies with long-term periods,
which are related to positive and negative superhumps, are in orange crosses
and blue triangles, respectively (as listed in Table~\ref{tab:cvprec}).  The
filled magenta circles denote the results from our study.  The gray area
denotes the possible region of disk precession period $P_\mathrm{prec}$ with
superhump period excess $\epsilon$ in $1 - 7\%$ range (positive superhump
($P_\mathrm{sh+}$)).  The iris hatched area represents the region of accretion
disk precession period $P_\mathrm{prec}$ with period excess of superhump
($\epsilon$) between $-0.5$\% and $-3.5$\% (negative superhump
($P_\mathrm{sh-}$)).  The zoomed-in view of the dashed box is in upper right
corner of the plot.  Only CT Boo and V825 Her are located in the possible
regions related to the superhump phenomenon.} \label{fig:ppplot}
\end{figure}

In general, the period excess for a positive superhump is in the range of $1 -
7\%$.  The period deficit for a negative superhump is about half of a excess of
the positive superhump \citep{Patterson1999}.  Therefore, the period ranges of
the disk precession can be derived with known $P_\mathrm{orb}$ and
$P_\mathrm{sh}$.  Figure~\ref{fig:ppplot} presents our results in a plot of
$P_\mathrm{orb}$ vs. $P_\mathrm{long}$.

In general, LMXBs are similar to CVs \citep{2002ASPC..261..223C}.  LMXBs have
low mass companions; therefore, some of them possibly satisfy the mass-ratio
criterion that introduces the disk precession.  The results of some long-term
variability studies on LMXBs \citep{2012MNRAS.420.1575K} and CVs (see
Table~\ref{tab:cvprec}) were gathered and included with our
Figure~\ref{fig:ppplot} results for comparison.  In Figure~\ref{fig:ppplot},
the gray area and iris hatched areas denote the possible regions of the
long-term periods in the positive ($P_\mathrm{sh+}$) and negative superhump
($P_\mathrm{sh-}$) systems, respectively.  In this study, only the long-term
periods of CT Boo and V825 Her are the possible precession periods (see
Section~\ref{sec:longterm} for more details). 

If the long-term period is the disk precession period, the period is confined
by its orbital period. For an extreme case of CV, say $P_\mathrm{orb} = 1$ day,
the maximum apsidal precession period and the maximum nodal precession period
are approximately 101 d and 199 d, respectively.  We conclude that if the
long-term periodicities are longer than 200 d, then the possibility of the
long-term period being the disk precession can be completely eliminated.

\subsection{Hierarchical Triple Star} \label{sec:mec:triple} 
Some of the close binaries, such as LMXBs and CVs, were reported in triple or
even multiple star systems.  \citet{2001ApJ...563..934C} indicated that the
superorbital modulation period $P = 171.033 \pm 0.326$ d is found in LMXB 4U
1820-30 with a limit of the period derivative $|\dot{P}| < 1.03\times 10^{-4}$
d d$^{-1}$. The luminosity modulation and its long-term stability suggest that
it is a hierarchical triple star system.  \citet{2011A&A...526A..53B}
discovered a third star of mass $M_3$ = 6.1 $\pm$ 0.5 $M_\mathrm{J}$ orbiting
the eclipsing binary DP Leo with orbital period $P_3$ = 28 $\pm$ 2 yrs.
\citet{2011MNRAS.416.2202P} studied the eclipsing polar UZ For and found that,
except for the binary orbital period $P_\mathrm{orb} = 0.088$ d, there are also
two long-term cyclic variations: $P_3$ = 16 $\pm$ 3 yrs and $P_4$ = 5.25 $\pm$
0.25 yrs. Moreover, they were interpreted as two giant exoplanets, with masses
of $M_3 = 6.3 \pm 1.5 M_\mathrm{J}$ ($M_J$ is the Jovian mass) and $M_4 = 7.7
\pm 1.2 M_\mathrm{J}$, being companions to the binary system.
\citet{2012A&A...538A.122C} detected a periodicity of $\sim 900$ d in the light
curves of FS Aur, and then used a numerical simulation of a third body with
sub-stellar mass orbiting around the binary to reproduce the long-term
modulation in the binary eccentricity variation.

A triple star is expected to form in high number density regions, such as the
core of globular clusters, which may aid the capture process by a CV system.
\citet{2008MNRAS.387..815T} performed an N-body simulation for investigating
the formation of triple systems in star clusters. Their results yielded that
the probability of forming triple stars in the center of a star cluster is only
two orders lower than that of binaries, which is comparably prominent.  A small
third body could significantly affect the eccentricity of the inner binary, and
then result in a mass-transfer-rate change of the CV system. This long-term
periodicity is known as Lidov-Kozai oscillation
\citep{1962P&SS....9..719L,1962AJ.....67..591K}.  The relation between the
outer third star and the inner binary can be expressed by the relation proposed
by \citet{1979A&A....77..145M}:
\begin{equation}
    P_\mathrm{long} =
    P_{1,2}\left(\frac{a_3}{a_{1,2}}\right)\frac{m_1+m_2}{m_3},
    \label{eq:lk01}
\end{equation}
where $P_\mathrm{long}$ is the long-term periodicity, $P_{1,2}$ is the inner binary
orbital period, $m_1$ and $m_2$ are the primary and the secondary masses, $m_3$ is the
mass of the third star, $a_{1,2}$ is the separation of the inner CV, and $a_3$ is
the separation between the CV and the third star.  An approximation can be
expressed as follows:
\begin{equation}
    P_\mathrm{long} \approx
    \alpha\frac{P_\mathrm{3}^2}{P_{1,2}}\frac{m_1+m_2+m_3}{m_3}(1-e_\mathrm{out})^{3/2},
    \label{eq:lk02}
\end{equation}
where $e_\mathrm{out}$ is the eccentricity between the third star and the inner
binary, and $\alpha$ is a dimensionless quantity from the three-body
Hamiltonian.  When the outer eccentricity is fixed, and there is no mass
transfer between the inner binary and outer star, then Equation~\ref{eq:lk02}
can be expressed as follows:
\begin{equation}
    P_\mathrm{long} = K\frac{P_\mathrm{3}^2}{P_{1,2}},
    \label{eq:lk03}
\end{equation}
where $K$ is a constant of order unity, and \citet{2001ApJ...563..934C} take $K
\simeq 1$ as an approximation (for different inclination angles, $K$ taken as a
factor of 2 smaller or larger is applicable).  For example, in 4U 1820-30,
$P_\mathrm{long} = 171.033$ d and $P_\mathrm{orb} = 685$ s.  Using
Equation~\ref{eq:lk03}, and assuming $K = 1$, this results in $P_3 = 1.18$ d,
which is considerably longer than the orbital period of the inner binary, and
so the hierarchical order of the triple system is reasonable.  Therefore, we
could conclude that the hierarchical triple star system is one of the possible
scenarios leading to long-term periodicities in CVs.

\subsection{Magnetic Variation of the Companion Star} \label{sec:mec:magcomp}
Most companions in CVs are late-type stars.  The convective zone in a late-type
star (G, K, or M type) yields a strong magnetic field through the well-known
dynamo process.  The strength of the magnetic field is related to the rotation
of the star. Under the same rotation speed, a star of the latter spectral type
would yield stronger magnetic field by the dynamo effect
\citep{1982ApJ...253..290D, 1982A&A...108..322R}.  Because of the stronger
magnetic field strength, the accretion process in the CV system is possibly
affected. 

Long-term modulations of orbital period variations have been found and studied
in some eclipsing binaries. \citet{1992ApJ...385..621A} proposed that the
torque generated from the magnetic variation of the companion may induce such a
long-term periodicity.  \citet{1996A&A...310..519M} studied the outburst
frequency of dwarf novae and found a cycle of 18 yrs in SS Aur and 7 yrs of SS
Cyg. They inferred that these cycles are related to the magnetic field
variation of the companion.  \citet{2008A&A...480..481B} discovered a 36 yr
periodic signal in the eclipsing cataclysmic variable HT Cas.  The periodicity
is likely to be the solar-type magnetic activity of the companion stars.
\citet{2001A&A...369..882A} observed cyclical variations in quiescence
magnitude and outburst intervals of 21 dwarf novae, one nova, and one NL star.
The probability density functions are peaked at 9.7 yrs for CVs, 7.9 yrs for
single main-sequence stars, and 8.6 yrs for all stars.

The distribution of the cycle lengths caused by the magnetic field variations
of late-type stars was studied by \citet{2016A&A...595A..12S}. Their results
reveal that the majority of cycle lengths are in the $2-13$ yr range.  For our
results in Table~\ref{tab:summary}, the longest period is $P_\mathrm{long} =
515.52 \pm 1.85$ d in V825 Her, which is marginal relative to the
aforementioned magnetic time scale. We cannot completely eliminate the
possibility of magnetic variation of the companion as a cause of long-term
periodicity in CVs. However, for our targets, the chance of magnetic variation
is not high according to the revealed period ranges of our targets.

\subsection{Other Possibilities}
In addition to the above mechanisms for long-term periodicities in CVs, the
superoutburst cycle of the SU UMa type of CV sometimes displays periodic-like
signals \citep[e.g.][]{2009JAVSO..37...80S}.  On the other hand,
\citet{2016MNRAS.463.1342S} found the periodic-like transition between the high
and low states in cataclysmic variable AM Her, and proposed the lifetime of the
active region in the companion may contribute to the variation of the accretion
rate.  In some rare cases, the long-term variation may also possibly be induced
by the precession of the jet in XBs \citep{2012MNRAS.420.1575K}, but the
existence of jets in CVs is still controversial. However,
\citet{2011MNRAS.418L.129K} first observed the radio emissions of V3885 Sgr,
which is a nova-like cataclysmic variable (NLCV), and the radio emission from
V3885 Sgr is regarded as synchrotron emissions from the jet.

\begin{table}
\centering
\caption{Long-term periodicities in CVs caused by superhumps \label{tab:cvprec}}
\begin{tabular}{crrrccl}
\hline
\hline
Name & $P_\mathrm{orb}$ & $P_\mathrm{sh}$ & $P_\mathrm{long}$ & Type(sh) &
Type & Reference \\
& (hr) & (hr) & (day) & && \\
\tableline
\decimals
AH Men    & 2.95 & 3.05 & 3.71  & $+$ & Nova   & \citet{1995PASP..107..657P} \\
DV UMa    & 2.06 & 2.14 & 2     & $+$ & DN     & \citet{2006JAVSO..35..132V} \\
CC Scl    & 1.38 & 1.44 & 1.6$^\dag$  & $+$ & Nova   & \citet{2012MNRAS.427.1004W} \\
LT Eri    & 4.08 & 3.98 & 5.3   & $-$ & CV     & \citet{2005PASA...22..105A} \\
LQ Peg    & 3.22 & 3.42 & 2.37  & $+$ & Nova   & \citet{2012NewA...17..453R} \\
MV Lyr    & 3.19 & 3.31 & 3.6   & $+$ & Nova   & \citet{1995PASP..107..545S} \\
NSV 1907  & 6.63 & 6.22 & 4.21  & $-$ & NL     & \citet{2017NewA...50...30H} \\
{\catcode`\&=11
\gdef\2010AnA...514A..30T{\citet{2010A&A...514A..30T}}}
PX And    & 3.41 & 3.51 & 4.43  & $+$ & CV     & \2010AnA...514A..30T \\
RR Pic    & 3.48 & 3.78 & 1.79  & $+$ & Nova   & \citet{2008MNRAS.389.1345S} \\
{\catcode`\&=11
\gdef\2001AnA...379..185S{\citet{2001A&A...379..185S}}}
TT Ari    & 3.30 & 3.57 & 1.82  & $+$ & Nova   & \2001AnA...379..185S \\
{\catcode`\&=11
\gdef\1994AnAS..107..219A{\citet{1994A&AS..107..219A}}}
TV Col    & 5.50 & 5.20 & 3.93  & $-$ & DQ Her & \1994AnAS..107..219A \\
UX UMa    & 4.72 & 4.48 & 3.68  & $-$ & NL     & \citet{2016MNRAS.457.1447D} \\
V1193 Ori & 3.43 & 3.26 & 2.98  & $-$ & Nova   & \citet{2005NewA...11..147A} \\
V2051 Oph & 1.50 & 1.54 & 52.5$^\dag$ & $+$ & DN     & \citet{2003MNRAS.338..165V} \\
V603 Aql  & 3.32 & 3.47 & 3.15$^\dag$ & $+$ & Nova   & \citet{2004AstL...30..615S} \\
WZ Sge    & 1.33 & 1.37 & 5.7   & $+$ & DN     & \citet{2002PASP..114..721P} \\
\tableline
\multicolumn{7}{p{15cm}}{ Note: $P_\mathrm{orb}$, $P_\mathrm{sh}$,
$P_\mathrm{long}$ are orbital, superhump and long-term disk precession periods,
respectively. Type(sh) is the type of superhump: ``$+$'' for positive
superhumps, ``$-$'' for negative superhumps.} \\
\multicolumn{7}{p{15cm}}{$^\dag$ average value of multiple periodicities}
\end{tabular}
\end{table}

\section{Long-term modulation for the individual CVs} \label{sec:longterm}
We have searched for long-term periodicities in approximately 100 CVs from the
PTF database (Section~\ref{sec:timing}).  It turns out that 10 of the targets
exhibit obvious periodic signals.  We folded the light curves with their best
periods, as shown in Figure~\ref{fig:flc1}, \ref{fig:flc2} and \ref{fig:flc3}.
The most convincing CVs that possess long-term periods ($P_\mathrm{long}$) are
discussed in this section.  The long-term periodicities discovered in these CVs
provide us with some probable implications for their formation mechanisms (see
Section~\ref{sec:mechanism}).  We discuss these CVs in the following
subsections:

\subsection{BK Lyncis (2MASS J09201119+3356423)}
BK Lyn (also PG0917+342), a NLCV, which was discovered by
\citet{1982PASP...94..560G}, has an orbital period of $P_\mathrm{orb}$ = 107.97
$\pm$ 0.07 min \citep{1996MNRAS.278..125R}.  \citet{2000MNRAS.314..826D} found
that its secondary is an M5V star from infrared spectroscopy.  The accretion
rate was estimated as $\dot{M} \sim$ 1 $\times 10^{-9} M_\odot/\mathrm{yr}$
assuming a WD mass of 1.2$~M_\odot$, or $\dot{M} \sim$ 1 $\times 10^{-8}
M_\odot/\mathrm{yr}$ for WD mass $M_\mathrm{WD}$ = 0.4$~M_\odot$
\citep{2009PASP..121..942Z}.  \citet{2012SASS...31....7K} studied 20-yr light
curves of BK Lyn and detected two sets of superhump signals.  One is 4.6\%
longer and the other is 3.0\% shorter than the $P_\mathrm{orb}$.  The two
superhumps were observed in different light curve stages.  They are possibly
due to the prograde apsidal precession and retrograde nodal precession of its
accretion disk.  The long-term period found in our study is $42.05 \pm 0.01$ d,
which is considerably longer than the long-term period derived from the
superhump periods found by \citet{2012SASS...31....7K}.  This suggests that the
periodicity is probably not from the precession of the accretion disk.  On the
other hand, if the long-term periodicity is driven by a third star orbiting the
CV system, then the orbital period of the third star will be $P_3 = 1.78$ d (as
calculated by Equation~\ref{eq:lk03}), which is much longer than the orbital
period of BK Lyn.

\subsection{CT Bo\"{o}tis}
CT Boo shows spectra similar to F-G stars, and no emission lines of hot
continuum was observed \citep{1994A&AS..107..503Z}.  If the spectrum is
dominated by its companion star, owing to the dynamo effect in the convective
layer of a late-type star, the magnetic field will be stronger than for other
spectral types of stars.  However, the long-term period is $P_\mathrm{long} =
7.91 \pm 1.05$ d, which is considerably shorter than the time scale of the
stellar magnetic variation (see Section~\ref{sec:mec:magcomp}).

Two possible orbital periods discovered by this study are $P_\mathrm{orb} =
230.72 \pm 1.78$ min and $P_\mathrm{orb} = 78.65 \pm 0.36$ min.  With the
positive superhump period excess between $1\%$ and $7\%$ and negative superhump
period excess between $-0.5\%$ and $-3.5\%$, we calculated the possible
accretion disk precession to be $2.45 - 31.88$ d for $P_\mathrm{orb} = 230.72$
min, and between 0.83 and 10.87 d for $P_\mathrm{orb} = 78.65$ min.  Therefore,
the long-term period found for CT Boo is in good agreement with the disk
precession period from the coupling of orbital and superhump periods.

\subsection{LU Camelopardalis: 2MASS J05581789+6753459}
The orbital period of LU Cam is $P_\mathrm{orb} = 0.1499686(4)$ d
\citep{2007PASP..119..494S} whereas its long-term modulation period is
$P_\mathrm{long} = 265.76 \pm 2.10$ d.  The possibility of disk precession can
be completely eliminated because it exceeds the period limit of 29.84 d for the
possible disk precession period.  If LU Cam is a hierarchical triple star
system, using Equation~\ref{eq:lk03} and assuming $K = 1$, the orbital period
of the outer third body is $P_3 \sim 6.20$ d, which is considerably longer than
the orbital period of the inner binary.  Therefore, the hierarchical triple
star is a possible scenario for the long-term periodicity in LU Cam.

\subsection{QZ Serpentis: SDSS J155654.47+210719.0}
QZ Ser is a dwarf nova system with an orbital period of 0.08316 d, and with a
K4 type star as its companion \citep{2002PASP..114.1117T}.  If there is
superhump-related disk precession, the period should be less than 16.54 d,
which is considerably shorter than the long-term period that we obtained
($P_\mathrm{long}$ = 277.72 $\pm$ 8.76 d).  If QZ Ser is a triple star system,
the orbital period of the third star would be $P_3 \sim 4.81$ d (through
Equation~\ref{eq:lk03}).

\subsection{V825 Herculis: 2MASS J17183699+4115511}
V825 Her, also named PG117+413 and 2MASS J17183699+4115511, has an orbital
period of $P_\mathrm{orb} = 4.94$ h discovered by \citet{1991BAAS...23.1463R}.
The upper limits of the long-term periods calculated for positive and negative
superhump systems are 20.79 d and 40.96 d, respectively.  There are two sets of
long-term periodicities in the power spectrum of V825 Her. One is
$P_\mathrm{long} = 9.24 \pm 0.05$ d, and the other one is considerably longer
$P_\mathrm{long} = 515.52 \pm 1.85$ d. These two periods exhibit convincing
variations in the folded light curves, as shown in Figure.~\ref{fig:flc2}.
$P_\mathrm{long} = 9.24$ d is plausibly explained as the disk precession
period; whereas, $P_\mathrm{long} = 515.52$ d is perhaps caused by the magnetic
variation of the companion star (see Section~\ref{sec:mec:magcomp}).

\subsection{V1007 Her: 1RXS J172405.7+411402}
V1007 Her, discovered by \citet{1998MNRAS.296..437G}, is a polar system with an
orbital period of $P_\mathrm{orb} = 119.93 \pm 0.0001 $ min. Because it is a
polar system, the long term modulation is impossibly caused by precession of
accretion disk.  If the long-term periodicity is caused by a third star, then
the orbital period of the third star ($P_3$) is about $3.77$ d, much longer
than the orbital period of the inner CV, and this becomes a possible mechanism
to explain the period of $\sim 170$ d modulation.

\subsection{UMa 01: 2MASS J09193569+5028261}
UMa 01 (also 2MASS J09193569+5028261) was identified by
\citet{2006ApJS..162...38A} as a CV.  The orbital period discovered in this
study is $P_\mathrm{orb} = 404.10 \pm 0.30 $ min.  If there is disk precession
in this system, then the precession period would range from $4.29$ to $55.84$
d.  However, the long-term period discovered in this study is $P_\mathrm{long}
= 246.84 \pm 0.81$ d, which is too large for disk precession.  On the other
hand, if the long-term periodicity is from a third star orbiting around the CV
system, then the orbital period $P_3$ would be $8.32$ d, much longer than the
orbital period of the inner CV, and so UMa 01 may be a hierarchical triple
system.

\subsection{Her 12: SDSS J155037.27+405440.0}
Her 12 (also SDSS J155037.27+405440.0) was identified as a CV by
\citet{2006ApJS..162...38A}.  We proposed two candidates of its orbital period
based on the LOT observations, $P_\mathrm{orb} = 75.62 \pm 0.008$ min and
$P_\mathrm{orb} = 173.65 \pm 1.57$ min.  The ranges for the disk precession
periods are $0.8$ to $10.45$ d and $1.84$ to $24$ d for $P_\mathrm{orb} =
75.62$ min and $173.65$ min, respectively.  The long-term modulation period
discovered in this work is $P_\mathrm{long} = 43.6 \pm 0.29$ d, exceeding both
ranges, and thus, this modulation is unlikely to be caused by accretion disk
precession.  On the other hand, if this source is a hierarchical triple system,
then the orbital period of the third companion is $1.51$ d or $2.29$ d for the
orbital periods of $75.62$ min or $173.65$ min, respectively. Both of the
third-body orbital periods are much larger than the corresponding orbital
periods of the CV. Therefore, this triple model is a possible mechanism to
explain the long-term variability of Her 12. 

\subsection{Coronae Borealis 06 \& VW Coronae Borealis}
CrB 06 (also 2MASS J15321369+3701046) and VW CrB (also USNO-B1.0 1231-00276740)
were identified as CVs in \citet{2006AJ....131..973S} and
\citet{2006ApJS..162...38A}, respectively.  Unfortunately, the orbital periods
of both CVs remain unknown. Therefore, at the current stage, we cannot
speculate on the possible mechanisms of long-term variability for these CVs.

\subsection{Summary}
We discussed several possible mechanisms to explain the long-term modulations
for the CVs that we found in this study.  Only the long-term periodicities in
CT Boo and V825 Her are possibly caused by the disk precession periods.  On the
other hand, for the other CVs included in this study, the hierarchical triple
systems are conceivable mechanisms for yielding long-term periodicities in CVs.
In addition, the magnetic variations of companions in CVs is yet another
possible mechanism to cause the long-term variations; nevertheless, the
tendency toward longer periods by this mechanism reduces its likelihood as a
plausible explanation for the long-term periodicities in this study.

\section{Conclusion and prospect} 
We presented our study on the long-term variations of CVs in this paper.  We
matched 344 CVs in the CV catalog of Downes with the PTF photometric database.
Approximately 100 CVs have more than 100 observations.  Among them, 10 of the
CVs exhibited long-term periodicities. The long-term periodicities may be
caused by one of the following mechanisms: disk precession, the presence of a
hierarchical triple-star system, magnetic variation of the companion star, jet
precession, and others. We discussed the likelihood of each possible mechanism
with respect to our matched CVs.  In addition, we found the possible orbital
periods for three of the CVs with the short cadence optical observations made
by LOT. The ranges of the orbital periods are typical in the orbital period
distribution.

With less sparse observations and longer observation time spans, the
uncertainties in the long-term periodic signals can be reduced. We look forward
to more sophisticated sample of long-term periodicities in CVs when ZTF and
other big synoptic surveys are in operation.  According to the current plan for
ZTF, the Galactic-plane survey will be one of the major parts of the project
that may considerably enhance the target number of CVs. We will certainly
benefit considerably from next-generation synoptic surveys.

\acknowledgements
M.T.-C.Y. is grateful to have the anonymous referee for the precious comments
and suggestions.  M.T.-C.Y. would like to thank the Ministry of Science and
Technology (MOST) of Taiwan for support through the grant MOST
104-2119-M-008-024 (TANGO Project, as the courtesy of W.H.Ip.) and NSC
102-2112-M-008-020-MY3.  PTF is a collaboration project in science among the
California Institute of Technology, Los Alamos National Laboratory, the
University of Wisconsin-Milwaukee, the Oskar Klein Center, the Weizmann
Institute of Science, the TANGO Program of the University System of Taiwan, and
the Kavli Institute for the Physics and Mathematics of the Universe.  CRTS was
supported by the NSF grants of AST-1313422, AST-1413600, and AST-1518308.





\end{document}